{}
{}

\documentclass[11pt,a4paper]{article}

\newif\ifpublic\publictrue

\ifpublic\else\usepackage{showkeys}\fi
\usepackage[bookmarks=true,bookmarksnumbered,hyperfigures=true,hidelinks]{hyperref}
\usepackage[nosort]{cite}
\usepackage[parsep]{collref}
\usepackage[english]{babel}
\usepackage[T1]{fontenc}
\usepackage[latin1]{inputenc}
\usepackage{amsfonts,amsbsy,dsfont,euscript,mathrsfs,fixmath}
\usepackage{amsmath,amssymb}
\usepackage{enumerate}
\usepackage{array}
\usepackage{delimset}

\def\showkeysrefformat#1{{\normalfont\tiny\ttfamily#1}}
\makeatletter
\def\SK@@ref#1>#2\SK@{{\@inlabelfalse\leavevmode\vbox to\z@{\vss\SK@refcolor\rlap{\vrule\raise .75em \hbox{\showkeysrefformat{#2}}}}}}
\makeatother

\usepackage[a4paper,text={160mm,247mm},centering]{geometry}
\allowdisplaybreaks[3]
\numberwithin{equation}{section}
\def\[{\begin{equation}\begin{aligned}}
\def\]{\end{aligned}\end{equation}}
\newcommand{\nn}{\nonumber}
\newcommand{\nln}{\nonumber\\}

\usepackage[font=small,labelfont=bf,width=0.85\textwidth]{caption}

\expandafter\def\expandafter\bfseries\expandafter{\bfseries\ifmmode\else\boldmath\fi}
\expandafter\def\expandafter\mdseries\expandafter{\mdseries\ifmmode\else\unboldmath\fi}
\expandafter\def\expandafter\normalfont\expandafter{\normalfont\ifmmode\else\unboldmath\fi}

\RequirePackage{verbatim}

\makeatletter
\newwrite\bibinl@out
\newenvironment{bibtex}[1][\jobname]{%
\immediate\openout\bibinl@out #1.bib%
\immediate\write\bibinl@out{\@percentchar generated from `\jobname' starting line \the\inputlineno^^J}%
\def\verbatim@processline{\immediate\write\bibinl@out{\the\verbatim@line}}%
\@bsphack\let\do\@makeother\dospecials\catcode`\^^M\active\verbatim@start%
}
{\immediate\closeout\bibinl@out\@esphack}
\makeatother

\let\barefrac=\frac
\renewcommand{\frac}[2]{\mathinner{\barefrac{#1}{#2}}}

\let\baresqrt=\sqrt
\makeatletter
\renewcommand{\sqrt}{\@ifnextchar[\@sqrt@space@a\@sqrt@space@b}
\def\@sqrt@space@a[#1]#2{\mathinner{\mathchoice{\mkern-3mu}{\mkern-3mu}{}{}\baresqrt[#1]{#2}}}
\def\@sqrt@space@b#1{\mathinner{\mathchoice{\mkern-3mu}{\mkern-3mu}{}{}\baresqrt{#1}}}
\makeatother

\makeatletter
\let\per@dot@old=\.
\def\.{\ifmmode\def\per@dot@sel{\mkern3mu}\else\def\per@dot@sel{\per@dot@old}\fi\per@dot@sel}
\makeatother

\let\barefootnote=\footnote
\renewcommand{\footnote}[1]{\barefootnote{#1\vspace{3pt}}}

\newcommand{\sfrac}[2]{{\textstyle\frac{#1}{#2}}}
\newcommand{\half}{\sfrac{1}{2}}

\newcommand{\vfrac}[2]{\ifmmode\mathinner{\textstyle^{#1}\!/\!_{#2}}\else$^{#1}\!/\!_{#2}$\fi}

\newcommand{\identity}{\mathds{1}}

\DeclareMathOperator{\diag}{diag}

\DeclareMathOperator{\STr}{STr}

\DeclareMathOperator{\sdet}{sdet}
\newcommand{\transpose}{t}
\newcommand{\supertranspose}{st}

\newcommand{\set}[1]{\{#1\}}

\newcommand{\Real}{\mathds{R}}
\newcommand{\Complex}{\mathds{C}}
\newcommand{\Integer}{\mathds{Z}}

\let\Re\relax\DeclareMathOperator{\Re}{Re}
\let\Im\relax\DeclareMathOperator{\Im}{Im}

\DeclareMathOperator{\arsinh}{arsinh}

\newcommand{\ind}[1]{{\scriptscriptstyle{#1}}}

\newcommand{\alg}[1]{\mathfrak{#1}}
\newcommand{\grp}[1]{\mathrm{#1}}

\DeclareMathOperator{\Lie}{Lie}

\DeclareMathOperator{\Ad}{Ad}

\newcommand{\dsum}{\oplus}

\newcommand{\com}[2]{[#1,#2]}
\newcommand{\anticom}[2]{\{#1,#2\}}
\newcommand{\mcom}[2]{[#1,#2\}}

\def\<{\big\langle}
\def\>{\big\rangle}
\newcommand\gen{\mathbb}

\newcommand{\geom}[1]{\mathrm{#1}}

\newcommand{\AdS}{\geom{AdS}}

\newcommand{\Sp}{\geom{S}}

\newcommand{\To}{\geom{T}}

\newcommand{\extder}{\mathrm{d}}
\newcommand{\intder}{\iota}

\newcommand{\Act}{S}

\def\wasyfamily{\fontencoding{U}\fontfamily{wasy}\selectfont}
\def\Circle{\mbox{\wasyfamily\char35}}
\newcommand\RRF{\mathcal{F}}
\newcommand\Nn{N}

\providecommand{\href}[2]{#2}

\makeatletter
\def\mr@ignsp#1 {\ifx\:#1\@empty\else #1\expandafter\mr@ignsp\fi}
\newcommand{\multiref}[1]{\begingroup%
\xdef\mr@no@sparg{\expandafter\mr@ignsp#1 \: }%
\def\mr@comma{}\def\mr@dash{-}%
\@for\mr@refs:=\mr@no@sparg\do{%
\ifx\mr@refs\mr@dash\def\mr@comma{}--\else%
\mr@comma\def\mr@comma{,}\ref{\mr@refs}\fi}%
\endgroup}
\renewcommand{\eqref}[1]{(\multiref{#1})}
\makeatother

\makeatletter
\newcommand{\namedref}[2]{\hyperref[#2]{#1~\ref*{#2}}}
\newcommand{\secref}{\@ifstar{\namedref{Section}}{\namedref{sec.}}}
\newcommand{\appref}{\@ifstar{\namedref{Appendix}}{\namedref{app.}}}
\newcommand{\tabref}{\@ifstar{\namedref{Table}}{\namedref{tab.}}}
\newcommand{\figref}{\@ifstar{\namedref{Figure}}{\namedref{fig.}}}
\makeatother

\let\oldbib=\thebibliography
\def\thebibliography{\phantomsection\addcontentsline{toc}{section}{\refname}\oldbib}

\let\oldtoc=\tableofcontents
\def\tableofcontents{\phantomsection\addcontentsline{toc}{section}{\contentsname}\oldtoc}

\providecommand{\hypersetup}[1]{}
\providecommand{\texorpdfstring}[2]{#1}
\hypersetup{plainpages=false}
\hypersetup{pdfpagemode=UseNone}
\hypersetup{bookmarksnumbered=true}
\hypersetup{pdfstartview=FitH}
\hypersetup{colorlinks=false}
\hypersetup{citebordercolor={1 1 1}}
\hypersetup{urlbordercolor={1 1 1}}
\hypersetup{linkbordercolor={1 1 1}}

\makeatletter
\let\@keywords\@empty
\let\@subject\@empty
\providecommand{\keywords}[1]{\gdef\@keywords{#1}}
\providecommand{\subject}[1]{\gdef\@subject{#1}}
\def\thetitle{\@title}
\def\theauthor{\@author}
\def\thesubject{\@subject}
\def\thedate{\@date}
\def\thekeywords{\@keywords}
\makeatother
\AtBeginDocument{
\hypersetup{pdftitle={\thetitle}}
\hypersetup{pdfauthor={\theauthor}}
\hypersetup{pdfsubject={\thesubject}}
\hypersetup{pdfkeywords={\thekeywords}}}

\newif\ifshownote
\shownotetrue

\ifpublic\shownotefalse\fi

\ifshownote

\ifpdf\else\RequirePackage[active]{srcltx}\fi
\RequirePackage{xcolor}
\RequirePackage{ulem}

\newcommand{\remark}[2][]{{%
\def\emph{\textsl}%
\def\tmparga{#1}%
\def\tmpargb{BH}\ifx\tmparga\tmpargb\normalfont\sffamily\hspace{1ex}\color[rgb]{0.5,0,0}\fi
\def\tmpargb{FS}\ifx\tmparga\tmpargb\normalfont\sffamily\hspace{1ex}\color[rgb]{0,0.5,0}\fi
\def\tmpargb{}\ifx\tmparga\tmpargb\color[rgb]{0,0.5,0.5}\else\normalfont\sffamily\hspace{1ex}\textbf{#1:}\fi
#2%
\ifx\tmparga\tmpargb\else\hspace{1ex}\fi}}

\else

\newcommand{\remark}[2][]{\ignorespaces\unskip}

\fi

\title{Supergravity backgrounds of the \texorpdfstring{$\eta$}{eta}-deformed \texorpdfstring{\\}{} \texorpdfstring{$\AdS_2 \times \Sp^2 \times \To^6$}{AdS2 x S2 x T6} and \texorpdfstring{$\AdS_5 \times \Sp^5$}{AdS5 x S5} superstrings}
\author{Ben~Hoare and Fiona~K.~Seibold}

\begin{document}

\pdfbookmark[1]{Title Page}{title}
\thispagestyle{empty}

\vspace*{2cm}
\begin{center}
\begingroup\Large\bfseries\thetitle\par\endgroup
\vspace{1cm}

\begingroup\theauthor\par\endgroup
\vspace{1cm}

\textit{
Institut f\"ur Theoretische Physik,\\
Eidgen\"ossische Technische Hochschule Z\"urich,\\
Wolfgang-Pauli-Strasse 27, 8093 Z\"urich, Switzerland}
\vspace{5mm}

\begingroup\ttfamily\small
\verb+{+bhoare,fseibold\verb+}+@itp.phys.ethz.ch\par
\endgroup
\vspace{5mm}

\vfill

\textbf{Abstract}\vspace{5mm}

\begin{minipage}{15cm}\small
We construct supergravity backgrounds for the integrable $\eta$-deformations of the $\AdS_2 \times \Sp^2 \times \To^6$ and $\AdS_5 \times \Sp^5$ superstring sigma models.
The $\eta$-deformation is governed by an R-matrix that solves the non-split modified classical Yang-Baxter equation on the superisometry algebra of the model.
Such R-matrices include those of Drinfel'd-Jimbo type, which are constructed from a Dynkin diagram and the associated Cartan-Weyl basis.
Drinfel'd-Jimbo R-matrices associated with inequivalent bases will typically lead to different deformed backgrounds.
For the two models under consideration we find that the unimodularity condition, implying that there is no Weyl anomaly, is satisfied if and only if all the simple roots are fermionic.
For $\AdS_2 \times \Sp^2 \times \To^6$ we construct backgrounds corresponding to the three Dynkin diagrams.
When all the simple roots are fermionic we find a supergravity background previously obtained by directly solving the supergravity equations.
For $\AdS_5 \times \Sp^5$ we construct a supergravity background corresponding to the Dynkin diagram with all fermionic simple roots.
\end{minipage}

\vspace*{2cm}

\end{center}

\newpage

\tableofcontents

\section{Introduction}

The semi-symmetric space sigma model of \cite{Metsaev:1998it,Berkovits:1999zq} describes the type II Green-Schwarz superstring on various $\AdS$ supergravity backgrounds, including $\AdS_2 \times \Sp^2 \times \To^6$ and $\AdS_5 \times \Sp^5$.
The model is integrable \cite{Bena:2003wd,Magro:2008dv,Vicedo:2010qd} and admits an integrable deformation, the $\eta$-deformation, otherwise known as a Yang-Baxter deformation \cite{Delduc:2013qra,Delduc:2014kha}.
This generalises the $\eta$-deformations of the principal chiral model \cite{Klimcik:2002zj,Klimcik:2008eq} and the symmetric space sigma model \cite{Delduc:2013fga}.
The deformation is governed by an R-matrix that solves the non-split modified classical Yang-Baxter equation on the superisometry algebra of the undeformed background, that is $\alg{psu}(1,1|2)$ for $\AdS_2 \times \Sp^2 \times \To^6$ and $\alg{psu}(2,2|4)$ for $\AdS_5 \times \Sp^5$.

The question of whether the deformed models also describe the type II Green-Schwarz superstring on a supergravity background has received considerable attention in recent years.
The metric and B-field of the $\eta$-deformed $\AdS_5 \times \Sp^5$ superstring were constructed in \cite{Arutyunov:2013ega} and the Ramond-Ramond fluxes in \cite{Arutyunov:2015qva} for a particular choice of R-matrix.
These background fields do not solve the type II supergravity equations.
It was later understood that they instead satisfy a set of generalised type II supergravity equations that depend on a background Killing vector \cite{Arutyunov:2015mqj,Wulff:2016tju}.

A general answer to when the background is expected to be a supergravity background was given in \cite{Borsato:2016ose}.
The R-matrix should satisfy the so-called unimodularity condition.
In this paper we consider a certain class of solutions to the non-split modified classical Yang-Baxter equation known as Drinfel'd-Jimbo R-matrices \cite{Drinfeld:1985rx,Jimbo:1985zk,Belavin:1984}.
This corresponds to a $q$-deformation of the superisometry algebra \cite{Delduc:2014kha,Delduc:2013fga,Delduc:2016ihq}.
Starting from a particular Dynkin diagram and Cartan-Weyl basis, the Drinfel'd-Jimbo R-matrix annihilates the Cartan generators and multiplies the positive and negative roots of the superisometry algebra by $-i$ and $+i$ respectively.
As the superisometry algebras are Lie superalgebras they can be described by inequivalent Dynkin diagrams.
The corresponding R-matrices can then lead to different deformations.

The R-matrix considered in \cite{Arutyunov:2015qva} corresponds to the distinguished Dynkin diagram
\[
\Circle-\Circle-\Circle-\otimes-\Circle-\Circle-\Circle ~,
\]
where $\Circle$ and $\otimes$ denote bosonic and fermionic roots respectively.
One can check that this R-matrix does not satisfy the unimodularity condition of \cite{Borsato:2016ose} in agreement with the fact that the background fields do not solve the type II supergravity equations.
The various methods in the literature used for constructing the $\eta$-deformed backgrounds, see, for example, \cite{Araujo:2018rbc} and \cite{Severa:2018pag,Demulder:2018lmj}, appear to correspond to considering the distinguished Dynkin diagram.
This is related to the fact that the distinguished Dynkin diagram has the Dynkin diagram associated to the bosonic subalgebra as a sub-Dynkin diagram.

It is worth recalling that also for non-compact Lie algebras there may be inequivalent R-matrices, which correspond to reordering the roots relative to the signature matrix.
For $\AdS_2$ the isometry algebra is $\alg{su}(1,1)$, which has only one non-split R-matrix.
However, for $\AdS_5$ the isometry algebra is $\alg{su}(2,2)$, which has three non-split R-matrices\cite{Delduc:2014kha}.
These give rise to different deformations of $\AdS_5$ that have been studied in \cite{Delduc:2014kha,Hoare:2016ibq,Araujo:2017enj}.
As these three R-matrices are different analytic continuations of the $\alg{su}(4)$ non-split R-matrix to $\alg{su}(2,2)$ it follows that the three metrics and B-fields are analytic continuations of each other \cite{Hoare:2014pna}.

\medskip

In this paper we investigate different Drinfel'd-Jimbo R-matrices and the corresponding $\eta$-deformed backgrounds.
For the two cases we consider, $\AdS_2 \times \Sp^2 \times \To^6$ and $\AdS_5 \times \Sp^5$, we find that the unimodularity condition of \cite{Borsato:2016ose} is satisfied if and only if all the simple roots of the corresponding Dynkin diagram are fermionic.
For $\AdS_2 \times \Sp^2 \times \To^6$ we consider the three inequivalent Dynkin diagrams of $\alg{psu}(1,1|2)$
\[
\Circle-\otimes-\Circle ~, \qquad \otimes-\Circle-\otimes ~, \qquad \otimes-\otimes-\otimes ~,
\]
and construct backgrounds corresponding to each one.
In the first two cases we find the same background, up to a shift in the B-field by a closed two-form, which agrees with the background of \cite{Arutyunov:2015mqj,Araujo:2018rbc} and solves the generalised type II supergravity equations.
In the final case, that is when all the simple roots are fermionic, we find the one-parameter background of \cite{Lunin:2014tsa}, constructed there by solving the type II supergravity equations directly, for a specific value of the parameter.
As a consequence of the bosonic roots not being simple, the Ramond-Ramond fluxes mix the $\AdS_2$ and $\Sp^2$ sectors in a non-trivial way.
In particular, they depend on a function that does not factorise into functions of the coordinates on $\AdS_2$ and functions of the coordinates on $\Sp^2$.
For $\AdS_5 \times \Sp^5$ we consider the Dynkin diagram of $\alg{psu}(2,2|4)$ with all fermionic simple roots
\[
\otimes-\otimes-\otimes-\otimes-\otimes-\otimes-\otimes ~.
\]
We choose an R-matrix that gives the metric and B-field of \cite{Arutyunov:2013ega} and derive the Ramond-Ramond fluxes and dilaton.
We verify that the background solves the type II supergravity equations.

The outline of this paper is as follows.
We start in \secref{sec:general} by summarising the key results from the literature that we use to construct the supergravity backgrounds.
In \secref{sec:ads2} and \secref{sec:ads5} we discuss the $\eta$-deformations of the $\AdS_2 \times \Sp^2 \times \To^6$ and $\AdS_5 \times \Sp^5$ superstrings respectively.
We conclude with a discussion of our results in \secref{sec:discussion}.
Our conventions for gamma matrices and superalgebras for the $\eta$-deformation of the $\AdS_2 \times \Sp^2 \times \To^6$ superstring are given in \appref{app:ads2}.
For the $\eta$-deformation of the $\AdS_5 \times \Sp^5$ superstring we follow the conventions of \cite{Arutyunov:2009ga,Arutyunov:2015qva}.

\section{General background}\label{sec:general}

In this section we summarise the key results that we use in our derivation of the (generalised) supergravity backgrounds for the $\eta$-deformations of the $\AdS_2 \times \Sp^2 \times \To^6$ and $\AdS_5 \times \Sp^5$ superstrings.

\paragraph{The \texorpdfstring{$\eta$}{eta}-deformation.}
The semi-symmetric space sigma model \cite{Metsaev:1998it,Berkovits:1999zq} is a sigma model on the supercoset
\[\frac{\grp{G}}{\grp{H}} ~, \]
where the basic Lie superalgebra $\alg{g} = \Lie(\grp{G})$ admits a $\Integer_4$ grading
\[
\alg{g} = \alg{g}_0 \oplus \alg{g}_1 \oplus \alg{g}_2 \oplus \alg{g}_3 ~,
\]
such that the grade 0 subalgebra $\alg{g}_0$ is identified with the Lie algebra of $\grp{H}$.
The subspaces $\alg{g}_0$ and $\alg{g}_2$ have even grading, while the subspaces $\alg{g}_1$ and $\alg{g}_3$ have odd grading.
We also introduce the projectors $P^{(i)}$ onto the subspaces $\alg{g}_i$.
Given a basis $\set{T_\ind{M}}$ of $\alg{g}$, we define
\[
\mathcal{K}_\ind{MN} = \STr[T_\ind{M}T_\ind{N}] ~,
\]
where $\STr$ denotes the supertrace, an ad-invariant and $\Integer_4$-invariant bilinear form on $\alg{g}$, which is symmetric (respectively antisymmetric) on the even (respectively odd) subspace of $\alg{g}$.
We also introduce the inverse of $\mathcal{K}_\ind{MN}$ through
\[
\mathcal{K}_\ind{MN} \widehat{\mathcal{K}}^\ind{NP} = \delta_\ind{M}^\ind{P} ~.
\]

The action of the $\eta$-deformed semi-symmetric space sigma model for the group-valued field $g \in \grp{G}$ is \cite{Delduc:2013qra,Delduc:2014kha}
\unskip\footnote{Given a particular matrix realisation of $\alg{g}$, $\STr$ is related to the usual matrix supertrace by a normalisation chosen such that the bosonic part of the action is given by
\[
\Act = - \frac{T}{2} \int \extder^2 \sigma (\gamma^{ij} G_{\mu \nu} -\epsilon^{ij} B_{\mu \nu}) \partial_i X^\mu \partial_j X^\nu
\]
where $G_{\mu \nu}$ is the target space metric and $B_{\mu \nu}$ the antisymmetric B-field.
The fields $X^\mu$ are the target space coordinates.
}
\[
\label{eq:etamodel}
\Act = - \frac{T}{4} (1-\eta^2) \int \extder^2 \sigma \, (\gamma^{ij} -\epsilon^{ij}) \STr[g^{-1} \partial_i g \, \hat{d} \frac{1}{1-\eta R_g \hat{d}} \, g^{-1} \partial_j g] ~,
\]
where $T$ is the overall coupling constant playing the role of the effective string tension, $\extder^2 \sigma = \extder \tau \extder \sigma$, $\gamma^{ij}$ is the Weyl invariant worldsheet metric with $\gamma^{\tau \tau}<0$ and $\epsilon^{ij}$ is the Levi-Civita symbol with $\epsilon^{\tau \sigma} = 1$.
The deformation parameter $\eta$ lies in the interval $(-1,1)$ and the operators $\hat{d}$ and $\hat{d}^t$ are defined in terms of the $\Integer_4$ projectors as
\[
\hat{d} = P^{(1)} + \frac{2}{1-\eta^2} P^{(2)} - P^{(3)}~, \qquad \hat{d}^\transpose = - P^{(1)} + \frac{2}{1-\eta^2} P^{(2)} + P^{(3)}~.
\]
The operator $R_g = \Ad_g^{-1} R \Ad_g^{\vphantom{-1}}$ acts on $X \in \alg{g}$ as $R_g(X)=g^{-1} R(gXg^{-1})g$.
The R-matrix is antisymmetric with respect to the supertrace
\[
\STr[R(X) Y]=-\STr[XR(Y)] ~,
\]
and solves the non-split modified classical Yang-Baxter equation
\[
\mcom{R(X)}{R(Y)}-R(\mcom{R(X)}{Y} + \mcom{X}{R(Y)}) = \mcom{X}{Y}~, \qquad X,Y \in \alg{g} ~.
\]
If we consider a purely bosonic group-valued field $g$, the action \eqref{eq:etamodel} reduces to the $\eta$-deformed symmetric space sigma model \cite{Delduc:2013fga}
\[
\Act = -\frac{T}{2} \int \extder^2 \sigma (\gamma^{ij} - \epsilon^{ij}) \STr[g^{-1} \partial_i g P^{(2)} \frac{1}{1-\kappa R_g P^{(2)}} g^{-1} \partial_j g] ~,
\]
where we have introduced the deformation parameter of \cite{Arutyunov:2013ega}.
\[
\kappa = \frac{2 \eta}{1-\eta^2} \in (-\infty,\infty) ~.
\]
The $\eta$-deformed semi-symmetric space sigma model has $q$-deformed symmetry \cite{Delduc:2014kha,Arutyunov:2013ega} with
\[
q = \exp \big(-\frac{\kappa}{T}\big) ~.
\]
such that $q$ is real.

\paragraph{Extracting the Ramond-Ramond fluxes.}
The background superfields of the $\eta$-deformed model (the supervielbein, NS-NS three-from, R-R bispinor, dilatino and gravitino field strength) can be extracted by comparing with the general form of the type II Green-Schwarz superstring \cite{Green:1983wt,Grisaru:1985fv,Tseytlin:1996hs,Cvetic:1999zs,Wulff:2013kga} following the procedure outlined in \cite{Borsato:2016ose}.
For completeness we summarise the important steps here.
Defining the operators
\[
O_+ = 1+ \eta R_g \hat{d}^\transpose ~, \qquad O_- = 1-\eta R_g \hat{d}~, \qquad M=O_-^{-1} O_+~,
\]
one observes that
\[
M^\transpose P^{(2)} M = P^{(2)}~,
\]
showing that $P^{(2)} M P^{(2)}$ implements a Lorentz transformation on the grade 2 subspace of $\alg{g}$.
Therefore, there exists an element $h \in \grp{H}$ such that
\[
\label{eq:Adhvector}
P^{(2)} M P^{(2)}= \Ad_h^{-1} P^{(2)}= P^{(2)} \Ad_h^{-1}~.
\]
Introducing $A_\pm = O_\pm^{-1} (g^{-1} \extder g)$ and defining the supervielbein as
\[
E^{(2)} = P^{(2)} A_+ ~, \qquad
E^{(1)} = \sqrt{1-\eta^2} \Ad^{\vphantom{-1}}_h P^{(1)} A_+~, \qquad
E^{(3)} = \sqrt{1-\eta^2} P^{(3)} A_-~,
\]
the action and the kappa symmetry variations take the standard Green-Schwarz form.
By calculating the superspace torsion and comparing the result with the general expression in \cite{Wulff:2016tju} one can obtain the background superfields.
In particular, the formula for the R-R bispinor is
\[
\label{eq:bispinorNUM}
\mathcal S^{1 \alpha 2 \beta} = 8 i \brk!{\Ad_h^{\vphantom{-1}}(1+\frac{2}{1-\eta^2} - 4 O_+^{-1})}^{1 \alpha }{}_{1 \gamma} \widehat{\mathcal{K}}^{1 \gamma 2 \beta }~,
\]
where the indices $\set{I = 1,2}$ and $\set{\alpha}$ combined run over the fermionic generators of $\alg{g}$ and the action of an operator $O$ on the basis $\set{T_\ind{M}}$ is given by
\[
O(T_\ind{M}) = T_\ind{N} O^\ind{N}{}_\ind{M} ~.
\]
This expression can then be compared with the familiar form of the R-R bispinor (written here for the R-R fluxes of type IIB supergravity in terms of $16 \times 16$ chiral gamma matrices)
\[
\label{eq:bispinorTH}
\mathcal{S} = -i \sigma_2 \gamma^a \RRF_{a} - \frac{1}{3 !} \sigma_1 \gamma^{abc} \mathcal{F}_{abc} - \frac{1}{2 \cdot 5!} i \sigma_2 \gamma^{abcde} \mathcal{F}_{abcde} ~,
\]
to extract the R-R fluxes.
For backgrounds with less than 32 supersymmetries (such as $\AdS_2 \times \Sp^2 \times \To^6$) the gamma matrices involve an additional projector to match the number of spinor indices.

Since we are only interested in the target space geometry and not its supergeometry it will be sufficient to take a purely bosonic group-valued field $g$.
Comparing \eqref{eq:bispinorNUM} and \eqref{eq:bispinorTH} gives the one-form $\RRF_1$, three-form $\RRF_3$ and five-form $\RRF_5$.
For standard supergravity backgrounds the R-R fluxes are then given by
\[
F_n = e^{-\Phi} \RRF_n ~,
\]
where the dilaton $\Phi$ is
\[
e^{-2\Phi} = e^{-2\Phi_0} \sdet(O_+)~.
\]
The R-R fluxes are defined in terms of the R-R potentials $C_n$ through
\[
F_n = \extder C_{n-1} + H \wedge C_{n-3} ~, \qquad H = \extder B ~,
\]
where $B$ is the B-field.
Henceforth, we will refer to both the forms $F_n$ and $\RRF_n$ as R-R fluxes.

In our conventions the Hodge star $\star$ acts on a $n$-form $A_n = \frac{1}{n!} A_{\mu_1 \dots \mu_n} \extder X^{\mu_1} \wedge \dots \wedge \extder X^{\mu_n}$ as
\[
(\star A_n)_{\mu_1 \dots \mu_{d-n}} = \frac{1}{n!} \sqrt{-G} \epsilon_{\mu_1 \dots \mu_{d-n} \nu_1 \dots \nu_n} A^{\nu_1 \dots \nu_n}~,
\]
where $G$ is the determinant of the metric.
The self-duality condition for the five-form reads $\RRF_5 = \star \RRF_5$.

\paragraph{The condition for Weyl invariance.}
It was shown in \cite{Borsato:2016ose} that for the $\eta$-deformation to be Weyl invariant, that is for the background fields to solve the type II supergravity equations, the unimodularity condition
\[\label{eq:unimodularity}
\widehat{\mathcal{K}}^\ind{MN} \STr[\mcom{T_\ind{M}}{R(T_\ind{N})} Z]=0 ~, \qquad \forall Z \in \alg{g} ~,
\]
should be satisfied.

Given a bosonic Lie algebra $\alg{f}$ and an R-matrix satisfying the modified classical Yang-Baxter equation on $\alg{f}$, the R-bracket
\[
\com{X}{Y}_R = \com{X}{R(Y)} + \com{R(X)}{Y} ~, \qquad X,Y \in \alg{f} ~,
\]
defines an alternative Lie bracket on $\alg{f}$ \cite{SemenovTianShansky:1983ik}.
We denote the structure constants of the R-bracket by $\tilde{f}_\ind{MN}{}^\ind{P}$.
Then, at least for semi-simple $\alg{f}$, the unimodularity condition \eqref{eq:unimodularity} is equivalent to the unimodularity of the Lie algebra generated by the R-bracket
\[
\sum_\ind{N} \tilde{f}_\ind{MN}{}^\ind{N} = 0 ~,
\]
that is the trace of the structure constants vanishes.

For Lie superalgebras the R-bracket
\[\label{eq:superrbracket}
\mcom{X}{Y}_R = \mcom{X}{R(Y)} + \mcom{R(X)}{Y} ~, \qquad X,Y \in \alg{g} ~,
\]
again defines an alternative Lie bracket whose structure constants we also denote $\tilde{f}_\ind{MN}{}^\ind{P}$.
For Lie superalgebras of the type that we are considering, the unimodularity condition \eqref{eq:unimodularity} implies that
\[\label{eq:superunimodularity}
\sum_\ind{N} (-1)^{[N]} \tilde{f}_\ind{MN}{}^\ind{N} = 0 ~,
\]
where we have $[N] = 0$ and $[N] = 1$ for bosonic and fermionic generators respectively.
That is the unimodularity condition for Lie superalgebras is equivalent to the supertrace of the structure constants vanishing.

For the non-split Drinfel'd-Jimbo R-matrices that we investigate in this paper, the R-bracket generates the positive Borel superalgebra (with two copies of the positive roots).
In app.~B of \cite{Hoare:2018ebg} the trace of the structure constants of the Borel superalgebra was computed for the three inequivalent Dynkin diagrams of $\alg{psl}(2|2;\Complex)$.
As we have seen above the relevant condition for the Weyl anomaly is the supertrace of the structure constants.
While the trace does not vanish for any Dynkin diagram, as we will discuss in \secref{sec:ads2}, the supertrace for the Dynkin diagram with all fermionic simple roots is zero.

\section{\texorpdfstring{$\eta$}{eta}-deformation of the \texorpdfstring{$\AdS_2 \times \Sp^2 \times \To^6$}{AdS2 x S2 x T6} superstring}\label{sec:ads2}

The first case we consider in detail is the $\eta$-deformation of the $\AdS_2 \times \Sp^2 \times \To^6$ superstring.
The $\Integer_4$ supercoset describing the curved part of the background is
\[\label{eq:supercosetads2}
\frac{\grp{PSU}(1,1|2)}{\grp{SO}(1,1) \times \grp{SO}(2)} ~.
\]
Considering the $\eta$-deformation of the semi-symmetric space sigma model on this supercoset, we analyse the different possible Drinfel'd-Jimbo R-matrices and find that those associated with the Dynkin diagram that has all fermionic simple roots satisfy the unimodularity condition \eqref{eq:unimodularity}.
For three R-matrices, associated with three inequivalent Dynkin diagrams, we construct the embedding of the $4$-dimensional background in $10$ dimensions following \cite{Sorokin:2011rr,Lunin:2014tsa} with the remaining compact dimensions given by a six-torus.
As expected, we find that for the unimodular R-matrices the background satisfies the standard type II supergravity equations, while for the non-unimodular R-matrices the generalised type II supergravity equations of \cite{Arutyunov:2015mqj,Wulff:2016tju} are satisfied.

\subsection{Choice of R-matrix}

The superisometry algebra of the $\AdS_2 \times \Sp^2$ semi-symmetric space is $\alg{psu}(1,1|2)$.
Its complexification $\alg{psl}(2|2; \Complex)$ admits three inequivalent Dynkin diagrams
\[\Circle-\otimes-\Circle ~, \qquad \otimes-\Circle-\otimes ~, \qquad \otimes-\otimes-\otimes ~.\]
Since the Drinfel'd-Jimbo R-matrix is defined through its action on the Cartan generators and the positive and negative roots, the three Dynkin diagrams give rise to three inequivalent R-matrices, and hence different $\eta$-deformations.

Following app.~D of \cite{Delduc:2014kha}, inequivalent R-matrices of $\alg{psu}(1,1|2)$ can be described in terms of permutations of 4 elements in the following way.
Working with the $4\times4$ supermatrix realisation of $(\alg{p})\alg{su}(1,1|2)$ given in \appref{app:ads2}, which has the property that the upper-left and lower-right $2\times2$ blocks have even grading and generate $\alg{su}(1,1)$ and $\alg{su}(2)$ respectively, we start from a certain reference R-matrix associated with the distinguished Dynkin diagram $\Circle-\otimes-\Circle$
\[\label{eq:refrmatrix}
\qquad R_{\text{0}}(M)_{ij} = - i \epsilon_{ij} M_{ij}~, \qquad
\epsilon = \begin{cases} \begin{aligned} + 1 \qquad & i < j ~, \\ 0 \qquad & i = j ~, \\ -1 \qquad & i > j ~.
\end{aligned} \end{cases}
\]
Now considering the permutation matrix $P_{ij} = \delta_{\mathds{P}(i)j}$ a new R-matrix can be constructed as
\unskip\footnote{From the form of $R_{\mathds{P}}$ it follows that an equivalent way of implementing the permutation and obtaining the different deformations is to keep the same R-matrix $R_0$ but act with the permutation matrix on the supermatrix realisation of $\alg{psu}(1,1|2)$.}
\[\label{eq:permutedrmatrix}
R_{\mathds{P}} = \Ad_P^{-1} R_0 \Ad_P^{\vphantom{-1}} ~.
\]

There are $4!$ possible permutations of 4 elements.
Of these the only ones of interest are those that preserve the ordering of $\set{1,2}$ and $\set{3,4}$.
This corresponds to considering R-matrices that have the same action on the $\alg{su}(1,1)$ and $\alg{su}(2)$ subalgebras respectively.
The permutation matrices that correspond to permuting $\set{1,2}$ or $\set{3,4}$ are related to elements of $\grp{SU}(1,1)$ and $\grp{SU}(2)$ respectively by multiples of the identity.
Therefore, due to the structure of the permuted R-matrix, these permutation matrices can be absorbed into a redefinition of the supergroup-valued field $g$.
We additionally only consider one of each pair of R-matrices related by the permutation
\[
\begin{pmatrix}
1 & 2 & 3 & 4
\\
3 & 4 & 1 & 2
\end{pmatrix} ~,
\]
since this amounts to simultaneously analytically continuing $\AdS_2 \to \Sp^2$ and $\Sp^2 \to \AdS_2$ and hence can be easily implemented directly on the (generalised) supergravity background.

We are left with 3 classes of permutations, each corresponding to a different Dynkin diagram.
The particular representatives of these classes that we consider are
\begin{align}\label{eq:r0}
\mathds{P}_0 =
\begin{pmatrix}
1 & 2 & 3 & 4
\\
1 & 2 & 3 & 4
\end{pmatrix} ~,
\qquad
R_{\mathds{P}_0} = R_0 ~, \qquad
\Circle-\otimes-\Circle~,
\\\label{eq:r1}
\mathds{P}_1 =
\begin{pmatrix}
1 & 2 & 3 & 4
\\
1 & 3 & 4 & 2
\end{pmatrix} ~,
\qquad
R_{\mathds{P}_1} = R_1 ~, \qquad
\otimes-\Circle-\otimes~,
\\\label{eq:r2}
\mathds{P}_2 =
\begin{pmatrix}
1 & 2 & 3 & 4
\\
1 & 3 & 2 & 4
\end{pmatrix} ~,
\qquad
R_{\mathds{P}_2} = R_2 ~, \qquad
\otimes-\otimes-\otimes~.
\end{align}
The R-matrix $R_2$, associated with the Dynkin diagram $\otimes-\otimes-\otimes$, satisfies the unimodularity condition \eqref{eq:unimodularity}, while the remaining two do not.

\subsection{The supergravity and generalised supergravity backgrounds}

To embed the $4$-dimensional backgrounds into $10$ dimensions we introduce the flat metric on the six-torus
\[
ds_{\To^6}^2 = dx^i dx^i ,
\]
where $i=4,\dots,9$, together with the holomorphic three-form $\Omega_3$ and K\"{a}hler form $J_2$
\[
\Omega_3 = \extder z^{1} \wedge \extder z^{2} \wedge \extder z^{3} ~, \qquad J_2 = \frac{i}{2} ( \extder \bar{z}^1 \wedge \extder z^1 + \extder \bar{z}^2 \wedge \extder z^2 + \extder \bar{z}^3 \wedge \extder z^3) ~,
\]
where we choose the complex coordinates as
\[z^1=x^4 - i x^5~, \qquad z^2 = x^6 - i x^7 ~, \qquad z^3 = x^8 - i x^9 ~.\]

We then take the following ansatz for the metric, B-field and R-R fluxes
\[
\extder s^2 & = \extder s_4^2 + \extder s_{\To^6}^2 ~, \qquad B = B_2 ,
\\
\RRF_1 &= 0~, \qquad \RRF_5 = \frac{1}{2} (1+\star) \widehat{\RRF}_2 \wedge \Re \Omega_3~,
\\
\RRF_3 &= \widehat{\RRF}_r \Re \Omega_3 + \widehat{\RRF}_i \Im \Omega_3 + \widehat{\RRF}_1 \wedge J_2 + \frac{1}{6} \star(\widehat{\RRF}_1 \wedge J_2 \wedge J_2 \wedge J_2)~,
\]
where $\extder s_4^2$ and $B_2$ are the metric and B-field of the $\eta$-deformation of $\AdS_2 \times \Sp^2$, $\widehat{\RRF}_r$ and $\widehat{\RRF}_i$ are zero-forms, $\widehat{\RRF}_1$ is a one-form and $\widehat{\RRF}_2$ is a two-form in $4$ dimensions.
Contracting the R-R fluxes with the gamma matrices gives the R-R bispinor
\[
\label{eq:bispinorAdS2}
\mathcal{S}^\ind{IJ} = \mathcal{P}_4 \Big[ - 4 \sigma_1^\ind{IJ} e^\Phi \Big( \widehat{\RRF}_r \gamma^{468} + \widehat{\RRF}_i \gamma^{579} - \widehat{\RRF}_a \gamma^a \gamma^{45} \Big) - \epsilon^\ind{IJ} e^\Phi \widehat{\RRF}_{ab} \gamma^{ab} \gamma^{468} \Big] \mathcal{P}_4 ~,
\]
where the projector
\[
\mathcal{P}_4 = \frac{1}{4} (\identity - \gamma^{4567} - \gamma^{4589} - \gamma^{6789}) ~,
\]
singles out a $4$-dimensional subspace of the original $16$-dimensional spinor space.
This makes it possible to compare the R-R bispinor \eqref{eq:bispinorAdS2} with the general formula \eqref{eq:bispinorNUM} since $\alg{psu}(1,1|2)$ indeed has four fermionic generators of grading 1 and four fermionic generators of grading 3.
Note that our conventions for gamma matrices and the superalgebra $\alg{psu}(1,1|2)$ are given in \appref{app:ads2}.

Using the parametrisations
\[
g=e^{- t P_0}e^{-\arsinh \rho \, P_1}e^{-\phi P_2}e^{-\arcsin r \, P_3}~, \qquad h=e^{\half \omega^{ab} J_{ab}}~,
\]
for the bosonic group-valued field $g \in \grp{G}$ and the element $h \in \grp{H}$ encoding the Lorentz transformation $\Ad_h$, we find that \eqref{eq:Adhvector} is indeed satisfied provided that
\[\omega^{01} = \arsinh \frac{2 \kappa \rho}{1- \kappa^2 \rho^2}~, \qquad \omega^{23} = - \arcsin \frac{2 \kappa r}{1+\kappa^2 r^2} ~.
\]
The metric and B-field are the same for all three choices of R-matrix and are \cite{Arutyunov:2013ega,Hoare:2014pna}
\[\label{eq:metbfield2}
\extder s^2 &= \frac{1}{1-\kappa^2\rho^2}\brk*{-(1+\rho^2) \extder t^2 + \frac{\extder \rho^2}{1+\rho^2}}
\\ & \qquad + \frac{1}{1+\kappa^2 r^2} \brk*{(1-r^2) \extder \phi^2 + \frac{\extder r^2}{1-r^2}} + \extder x^i \extder x^i~, \\
B &= - \frac{\kappa \rho}{1-\kappa^2\rho^2} \extder t \wedge \extder \rho - \frac{\kappa r}{1+\kappa^2 r^2} \extder \phi \wedge \extder r~.
\]
Let us recall that the $\eta$-deformation of $\Sp^2$ is equivalent to the deformed model of \cite{Fateev:1992tk} up to the B-field, which is a closed two-form.

For the R-matrix $R_0$ \eqref{eq:r0} corresponding to the distinguished Dynkin diagram we find the following R-R fluxes
\[\label{eq:back0}
\RRF_3 &= - N \brk*{ \kappa \rho \Re \Omega_3 + \kappa r \Im \Omega_3 } ~, \\
\RRF_5 &= - N \brk*{ \frac{1}{1-\kappa^2\rho^2} \extder t \wedge \extder \rho + \frac{\kappa^2 \rho r}{1+\kappa^2 r^2} \extder \phi \wedge \extder r } \wedge \Re \Omega_3
\\ & \qquad - N \brk*{\frac{\kappa^2 \rho r}{1-\kappa^2\rho^2} \extder t \wedge \extder \rho - \frac{1}{1+\kappa^2 r^2} \extder \phi \wedge \extder r } \wedge \Im \Omega_3 ~,
\\ N & = \frac{\sqrt{1+ \kappa^2}}{\sqrt{1-\kappa^2 \rho^2}\sqrt{1+\kappa^2 r^2}} ~,
\]
while for the R-matrix $R_1$ \eqref{eq:r1} we find
\[\label{eq:back1}
\RRF_3 &= N \brk*{\kappa \rho \Re \Omega_3 - \kappa r \Im \Omega_3} ~, \\
\RRF_5 &= - N \brk*{\frac{1}{1-\kappa^2\rho^2} \extder t \wedge \extder \rho - \frac{\kappa^2 \rho r}{1+\kappa^2 r^2} \extder \phi \wedge \extder r} \wedge \Re \Omega_3
\\ & \qquad + N \brk*{\frac{\kappa^2 \rho r}{1-\kappa^2\rho^2} \extder t \wedge \extder \rho + \frac{1}{1+\kappa^2 r^2} \extder \phi \wedge \extder r } \wedge \Im \Omega_3 ~,
\\ N & = \frac{\sqrt{1+ \kappa^2}}{\sqrt{1-\kappa^2 \rho^2}\sqrt{1+\kappa^2 r^2}} ~.
\]
These two backgrounds can be related by changing the sign of $t$ and $\rho$ and shifting the B-field by closed two-form.
Therefore, from the perspective of (generalised) supergravity, they are equivalent.
Furthermore, they are equivalent to the background given in app.~F of \cite{Arutyunov:2015mqj}, which solves the generalised type IIB supergravity equations with a certain choice of the Killing vector and generalised dilaton one-form.

The shift in the B-field depends only on the $\AdS_2$ sector.
Even though this shift is by a closed two-form and hence the Green-Schwarz sigma models on the backgrounds \eqref{eq:metbfield2}, \eqref{eq:back0} and \eqref{eq:metbfield2}, \eqref{eq:back1} agree up to a total derivative, this observation still appears to have an algebraic interpretation.
For the R-matrix $R_1$ the roots of $\alg{su}(1,1)$ are not simple roots, while the roots of $\alg{su}(2)$ are.
Therefore, the corresponding sigma model will exhibit the Poisson-Lie symmetry associated with the $\alg{su}(2)$ subalgebra exactly, that is, not up to total derivatives.
However, this will not necessarily be the case for the $\alg{su}(1,1)$ subalgebra.
Indeed, considering the Poisson-Lie symmetry associated with the full superalgebra $\alg{psu}(1,1|2)$, the deformed $\alg{su}(1,1)$ symmetry should involve the fermionic charges.

For the R-matrix $R_2$ \eqref{eq:r2} we find the following R-R fluxes
\[\label{eq:back2}
\RRF_3 &= - N \brk*{\kappa r(1+\kappa^2 r^2) \extder \rho + \kappa \rho(1-\kappa^2 \rho^2) \extder r } \wedge J_2 \\
& \qquad + N \brk*{ \kappa \rho(1-r^2) \extder \rho - \kappa r (1+\rho^2) \extder r } \wedge \extder t \wedge \extder \phi ~, \\
\RRF_5 &= N \brk*{(1+\kappa^2 r^2) \extder \rho - \kappa^2 \rho r (1+\rho^2) \extder r } \wedge \extder t \wedge \Re \Omega_3 \\
& \qquad - N \brk*{ \kappa^2 \rho r (1-r^2) \extder \rho + (1-\kappa^2 \rho^2) \extder r } \wedge \extder \phi \wedge \Im \Omega_3 ~,
\\ N & = \frac{\sqrt{1+\kappa^2}}{\sqrt{1-\kappa^2 \rho^2}\sqrt{1+\kappa^2 r^2}} \frac{1}{1-\kappa^2(\rho^2-r^2-\rho^2 r^2)} ~.
\]
As expected, this background satisfies the type IIB supergravity equations with the dilaton
\[
e^{-2\Phi} = e^{-2\Phi_0} \frac{(1-\kappa^2 \rho^2)(1+\kappa^2 r^2)}{1-\kappa^2(\rho^2-r^2-\rho^2 r^2)} ~,
\]
and $\epsilon_{t\rho\phi r456789}=+1$.
The R-R potentials are
\[\begin{gathered}\label{eq:rrpotentials}
C_2 = -\frac{e^{-\Phi_0} \sqrt{1+\kappa^2}}{\sqrt{1-\kappa^2(\rho^2-r^2-\rho^2 r^2)}}(\kappa \rho r J_2 - \kappa^{-1} \extder t\wedge \extder\phi) ~,
\\
C_4 = \frac{e^{-\Phi_0} \sqrt{1+\kappa^2}}{\sqrt{1-\kappa^2(\rho^2-r^2-\rho^2 r^2)}}(\rho \, \extder t \wedge \Re \Omega_3 - r \, \extder \phi \wedge \Im \Omega_3) ~.
\end{gathered}\]
In fact, this background is known.
In \cite{Lunin:2014tsa} a one-parameter, denoted $a$, family of backgrounds supporting the metric and B-field of the $\eta$-deformed $\AdS_2 \times \Sp^2$ superstring was constructed.
It transpires that the background \eqref{eq:back2} corresponds to the point $a=1$.

The one-parameter dilaton of \cite{Lunin:2014tsa} is
\[\label{eq:lrt}
e^{-2\Phi} = e^{-2\Phi_0}\frac{(1-\kappa^2 \rho^2)(1+\kappa^2 r^2)}{1+ \kappa^2(a^2(r^2-\rho^2) + r^2\rho^2) - 2\kappa \sqrt{1-a^2}\sqrt{1+a^2\kappa^2} r\rho} .
\]
Taking $a \in \Real_{\geq 0}$, we note that this dilaton is an even function of $\kappa$ only for $a=1$.
It would be interesting to understand if there is an R-matrix that gives the one-parameter background of \cite{Lunin:2014tsa}, or if this is only the case for $a=1$.

For the R-matrix $R_2$ neither the roots of $\alg{su}(1,1)$ nor $\alg{su}(2)$ are simple roots.
Therefore, this model is not expected to exhibit Poisson-Lie symmetry associated with these bosonic subalgebras.
Indeed, considering the Poisson-Lie symmetry associated with the full superalgebra $\alg{psu}(1,1|2)$, the deformed $\alg{su}(1,1)$ and $\alg{su}(2)$ symmetries should involve the fermionic charges.

\subsection{Limits}

To conclude this section we briefly discuss three interesting limits of the supergravity background \eqref{eq:metbfield2}, \eqref{eq:back2}.
These are the plane-wave, maximal deformation ($\kappa \to \infty$) and Pohlmeyer ($\kappa \to i$) limits.
The latter two limits were discussed in \cite{Lunin:2014tsa}.

The plane-wave limit \cite{Berenstein:2002jq,Blau:2002dy} is reached by first setting
\[
\label{eq:limpp2}
t = \mu x^+ + \frac{x^-}{\mu L^2}~, \qquad \phi = \mu x^+ - \frac{x^-}{\mu L^2}~,
\]
and rescaling
\[
\label{eq:limpp1}
\rho \rightarrow \frac{\rho}{L}~, \qquad r \rightarrow \frac{r}{L}~, \qquad T \rightarrow L^2 T~,
\]
where $T$ is the effective string tension.
Also rescaling $x^i \to L^{-1} x^i$, we then send $L \rightarrow \infty$ keeping $\mu$, $x^\pm$, $\rho$, $r$ and $x^i$ finite.
Recalling that the R-R potentials scale with the tension as $C_0 \sim T^0$, $C_2 \sim T^1$ and $C_4 \sim T^2$, in the limit $L \rightarrow \infty$ the two-form $C_2$ is closed, while the metric, B-field, dilaton and four-form $C_4$ are
\[
\label{eq:pp2}
\extder s^2 &= -4 \extder x^- \extder x^+ - \mu^2 (1+\kappa^2)(\rho^2 + r^2) \, (\extder x^+)^2 + \extder \rho^2 + \extder r^2 + \extder x^i \extder x^i~, \\
B &= - \mu \kappa ( \rho \, \extder x^+ \wedge \extder \rho + r \, \extder x^+ \wedge \extder r)~, \qquad e^{-2 \Phi} = e^{-2 \Phi_0}~, \\
C_4 &= \mu e^{-\Phi_0} \sqrt{1+\kappa^2} ( \rho \, \extder x^+ \wedge \Re \Omega_3 - r \, \extder x^+ \wedge \Im \Omega_3 )~.
\]
Further taking $\mu \to 0$ gives flat space with vanishing NS-NS and R-R fluxes.
To reach the flat space background directly we can rescale $t \to L^{-1} t$ and $\phi \to L^{-1} \phi$ before taking $L \to \infty$, rather than setting \eqref{eq:limpp2}.
Both the generalised supergravity backgrounds \eqref{eq:metbfield2}, \eqref{eq:back0} and \eqref{eq:metbfield2}, \eqref{eq:back1} also admit the same plane-wave and flat space limits.

The maximal deformation limit \cite{Arutyunov:2014cra,Pachol:2015mfa} is given by first rescaling
\[\label{eq:limmaxdef}
t \to \frac{t}{\kappa}~, \qquad \rho \to \frac{\rho}{\kappa}~, \qquad
\phi \to \frac{\phi}{\kappa}~, \qquad r \to \frac{r}{\kappa}~, \qquad T\to \kappa^2 T ~.
\]
We also rescale $x^i \to \kappa^{-1} x^i$ and then take $\kappa \to \infty$.
This limit corresponds to a contraction of the $q$-deformed symmetry algebra \cite{Pachol:2015mfa}.
In this limit we find the following supergravity background
\[\label{eq:maxdef2}
\extder s^2 &= \frac{1}{1-\rho^2}\brk*{-\extder t^2 + \extder \rho^2}
+ \frac{1}{1+r^2} \brk*{\extder \phi^2 + \extder r^2} + \extder x^i \extder x^i~, \\
B &= - \frac{\rho}{1-\rho^2} \extder t \wedge \extder \rho - \frac{r}{1+ r^2} \extder \phi \wedge \extder r~,
\qquad
e^{-2\Phi} = e^{-2\Phi_0} \frac{(1-\rho^2)(1+r^2)}{1-\rho^2+r^2} ~,
\\
C_2 &= - \frac{e^{-\Phi_0}}{\sqrt{1-\rho^2+r^2}}(\rho r J_2 - \extder t \wedge \extder \phi) ~,
\\
C_4 &= \frac{e^{-\Phi_0}}{\sqrt{1-\rho^2+r^2}}(\rho \, \extder t \wedge \Re \Omega_3 - r \, \extder \phi \wedge \Im \Omega_3) ~,
\]
which does not match the mirror $\AdS_2 \times \Sp^2 \times \To^6$ supergravity background of \cite{Arutyunov:2014cra,Arutyunov:2014jfa}.
Note that in the maximal deformation limit the generalised supergravity backgrounds \eqref{eq:metbfield2}, \eqref{eq:back0} and \eqref{eq:metbfield2} \eqref{eq:back1} remain generalised supergravity backgrounds and hence are different to \eqref{eq:maxdef2}.
Further rescaling
\[\label{eq:flatlimit}
t \to \frac{t}{L} ~, \qquad \rho \to \frac{\rho}{L} ~, \qquad
\phi \to \frac{\phi}{L} ~, \qquad r \to \frac{r}{L} ~, \qquad
T \to L^2 T ~,
\]
together with $x_i \to L^{-1} x_i$, in the limit $L \to \infty$ we recover flat space with vanishing NS-NS and R-R fluxes.
Indeed, the metric and B-field of the background \eqref{eq:maxdef2} describe an integrable deformation of flat space with $\kappa$-deformed $\alg{iso}(1,1) \dsum \alg{iso}(2)$ symmetry \cite{Borowiec:2015wua,Pachol:2015mfa}.

The Pohlmeyer limit \cite{Hoare:2014pna} is given by setting
\[\label{eq:limpohlmeyer}
t & = \frac{\mu x^+}{\epsilon} + \frac{\epsilon x^-}{\mu} ~, \qquad & \phi & = \frac{\mu x^+}{\epsilon} - \frac{\epsilon x^-}{\mu} ~,
\qquad \kappa = i \sqrt{1-\epsilon^2} ~,
\\
\rho & = \tan \alpha ~, \qquad & r & = \tanh \beta ~,
\]
and then taking $\epsilon \to 0^+$.
In this limit the B-field has a divergent part that is a closed two-form and no finite part.
Furthermore, we find that the two-form $C_2$ vanishes.
The remaining background fields give the following simple pp-wave supergravity background
\[\label{eq:pohlmeyer2}
\extder s^2 &= -4 \extder x^- \extder x^+ - \mu^2 \brk*{\sin^2\alpha + \sinh^2\beta } \extder x^+{}^2 + \extder \alpha^2 + \extder \beta^2
+ \extder x^i \extder x^i~,
\quad \, e^{-2\Phi} = e^{-2\Phi_0} ~, \\
C_4 & = \mu e^{-\Phi_0} ( \sin\alpha\cosh\beta \, \extder x^+ \wedge \Re \Omega_3
- \cos\alpha\sinh\beta \, \extder x^+ \wedge \Im \Omega_3 ) ~.
\]
Even though it involves taking $\kappa$ to be imaginary, in the Pohlmeyer limit the background becomes real.
Interestingly, the same background arises in the Pohlmeyer limit of the generalised supergravity backgrounds \eqref{eq:metbfield2}, \eqref{eq:back0} and \eqref{eq:metbfield2}, \eqref{eq:back1} \cite{Hoare:2015gda}.
In all three cases, taking the limit $\kappa \to i$ without rescaling the coordinates $t$ and $\phi$ gives \eqref{eq:pohlmeyer2} with $\mu = 0$, that is flat space.
As shown in \cite{Hoare:2014pna} the light-cone gauge-fixing ($x^+ = \tau$) of the pp-wave background \eqref{eq:pohlmeyer2} leads to the Pohlmeyer-reduced theory of the $\AdS_2 \times \Sp^2$ superstring \cite{Grigoriev:2007bu}, which is equivalent to the $\mathcal{N} = 2$ supersymmetric sine-Gordon model.

\section{\texorpdfstring{$\eta$}{eta}-deformation of the \texorpdfstring{$\AdS_5 \times \Sp^5$ superstring}{AdS5 x S5}}\label{sec:ads5}

The second case we consider is the $\eta$-deformation of the $\AdS_5 \times \Sp^5$ superstring.
The $\Integer_4$ supercoset describing this background is
\[\label{eq:supercosetads5}
\frac{\grp{PSU}(2,2|4)}{\grp{Sp}(1,1) \times \grp{Sp}(2)} ~.
\]
Considering the $\eta$-deformations of the semi-symmetric space sigma model on the supercoset \eqref{eq:supercosetads5}, we analyse the different possible Drinfel'd-Jimbo R-matrices and find that only those associated with the Dynkin diagram that has all fermionic simple roots satisfy the unimodularity condition \eqref{eq:unimodularity}.
Observing that the supergravity backgrounds corresponding to different R-matrices associated with this Dynkin diagram should be related by analytic continuation, we pick a particular representative and construct the $10$-dimensional background in this case.
As expected, we find that the background satisfies the type IIB supergravity equations.

\subsection{Choice of R-matrix}

The superisometry algebra of the $\AdS_5 \times \Sp^5$ semi-symmetric space is $\alg{psu}(2,2|4)$.
Its complexification $\alg{psl}(4|4;\Complex)$ admits 35 inequivalent Dynkin diagrams.
Note that, for our purposes, two Dynkin diagrams that are related by a $\Integer_2$ reflection in the central node, but are not identical, are considered inequivalent.

Following app.~D of \cite{Delduc:2014kha}, inequivalent R-matrices of $\alg{psu}(2,2|4)$ can be described in terms of permutations of 8 elements in the following way.
Working with the $8 \times 8$ supermatrix realisation of $\alg{p}\alg{su}(2,2|4)$ of \cite{Arutyunov:2009ga,Arutyunov:2015qva}, which has the property that the upper-left and lower-right $4 \times 4$ blocks have even grading and generate $\alg{su}(2,2)$ and $\alg{su}(4)$ respectively, we start from the reference R-matrix \eqref{eq:refrmatrix} associated with the distinguished Dynkin diagram
\[\Circle-\Circle-\Circle-\otimes-\Circle-\Circle-\Circle~.\]
As for the $\AdS_2 \times \Sp^2 \times \To^6$ case, new R-matrices can be constructed using a permutation matrix as in equation \eqref{eq:permutedrmatrix}.

Of the $8!$ elements of the permutation group $\grp{S}_8$ we only consider those that preserve the ordering of $\set{1,2,3,4}$ and $\set{5,6,7,8}$.
This corresponds to considering R-matrices that have the same action on the $\alg{su}(2,2)$ and $\alg{su}(4)$ subalgebras respectively.
In \cite{Delduc:2014kha} it was shown that permutations reordering $\set{5,6,7,8}$ all lead to equivalent R-matrices as the permutation matrix can be absorbed into a redefinition of the supergroup-valued field $g$.
This is the statement that there is only a single inequivalent R-matrix for $\alg{su}(4)$.

On the other hand, permutations reordering $\set{1,2,3,4}$ lead to three inequivalent R-matrices \cite{Delduc:2014kha}.
A particular choice of the three permutations is
\[\label{eq:su22permutations}
\begin{pmatrix}
1 & 2 & 3 & 4
\\
1 & 2 & 3 & 4
\end{pmatrix} ~, \qquad
\begin{pmatrix}
1 & 2 & 3 & 4
\\
1 & 3 & 2 & 4
\end{pmatrix} ~, \qquad
\begin{pmatrix}
1 & 2 & 3 & 4
\\
1 & 3 & 4 & 2
\end{pmatrix} ~.
\]
The resulting R-matrices are the three inequivalent non-split Drinfel'd-Jimbo R-matrices of $\alg{su}(2,2)$ and give rise to different deformations of $\AdS_5$.
However, as shown in \cite{Hoare:2016ibq}, these are related to each other by analytic continuation.
As this analytic continuation can be implemented directly on the (generalised) supergravity background we restrict to the ordering of $\set{1,2,3,4}$ that corresponds to the R-matrix of $\alg{su}(2,2)$ used in \cite{Arutyunov:2013ega,Arutyunov:2015qva}.

We additionally only consider one of each pair of R-matrices related by the permutation
\[
\begin{pmatrix}
1 & 2 & 3 & 4 & 5 & 6 & 7 & 8
\\
5 & 6 & 7 & 8 & 1 & 2 & 3 & 4
\end{pmatrix}~,
\]
as this amounts to simultaneously analytically continuing $\AdS_5 \to \Sp^5$ and $\Sp^5 \to \AdS_5$ and hence can also be implemented directly on the (generalised) supergravity background.

We are left with 35 classes of permutations, each corresponding to a different Dynkin diagram.
Of all the associated R-matrices only one satisfies the unimodularity condition \eqref{eq:unimodularity}.
This R-matrix is related to the reference R-matrix \eqref{eq:refrmatrix} by the permutation
\[
\mathds{P} =
\begin{pmatrix}
1 & 2 & 3 & 4 & 5 & 6 & 7 & 8
\\
1 & 5 & 2 & 6 & 3 & 7 & 4 & 8
\end{pmatrix}~,
\]
and hence is associated with the Dynkin diagram that has all fermionic simple roots
\[
\otimes-\otimes-\otimes-\otimes-\otimes-\otimes-\otimes ~.
\]
The explicit action of the R-matrix is given by
\[\label{eq:rmat5}
R(M)_{ij} = - i \epsilon_{ij} M_{ij}~,
\qquad \epsilon = \begin{pmatrix}
0 & +1 & +1 & +1 & +1 & +1 & +1 & +1 \\
-1 & 0 & +1 & +1 & -1 & +1 & +1 & +1 \\
-1 & -1 & 0 & +1 & -1 & -1 & +1 & +1 \\
-1 & -1 & -1 & 0 & -1 & -1 & -1 & +1 \\
-1 & +1 & +1 & +1 & 0 & +1 & +1 & +1 \\
-1 & -1 & +1 & +1 & -1 & 0 & +1 & +1 \\
-1 & -1 & -1 & +1 & -1 & -1 & 0 & +1 \\
-1 & -1 & -1 & -1 & -1 & -1 & -1 & 0 \\
\end{pmatrix}~.
\]

\subsection{The supergravity background}

To extract the background fields of the $\eta$-deformed $\AdS_5 \times \Sp^5$ superstring for the R-matrix \eqref{eq:rmat5} we follow \cite{Borsato:2016ose} using the same parametrisation for the group-valued field $g \in \grp{PSU}(2,2|4)$ as in \cite{Arutyunov:2013ega,Arutyunov:2015qva}.
For convenience we introduce the variables $x=\sin \zeta$ and $w=\sin \xi$.
The metric and B-field are \cite{Arutyunov:2013ega}
\[\label{eq:back5}
\extder s^2 &= \frac{1}{1-\kappa^2\rho^2}\brk*{-(1+\rho^2) \extder t^2 + \frac{\extder \rho^2}{1+\rho^2}} + \frac{\rho^2}{1+\kappa^2\rho^4 x^2} \brk*{ (1-x^2) \extder \psi_1^2+ \frac{\extder x^2 }{1-x^2} } + \rho^2 x^2 \extder \psi_2^2 \\
&\quad + \frac{1}{1+\kappa^2r^2}\brk*{(1-r^2) \extder \phi^2 + \frac{\extder r^2}{1-r^2}} + \frac{r^2}{1+\kappa^2r^4 w^2} \brk*{ (1-w^2) \extder \phi_1^2+ \frac{\extder w^2 }{1-w^2} } + r^2 w^2 \extder \phi_2^2~, \\
B &= \frac{\kappa \rho}{1-\kappa^2 \rho^2} \extder t \wedge \extder \rho + \frac{\kappa \rho^4 x}{1+\kappa^2 \rho^4 x^2} \extder \psi_1 \wedge \extder x + \frac{\kappa r}{1+\kappa^2 r^2} \extder \phi \wedge \extder r -\frac{\kappa r^4 w}{1+\kappa^2 r^4 w^2} \extder \phi_1 \wedge \extder w ~.
\]
For the R-R fluxes we find that the one-form $F_1$ vanishes, while the three-form and five-form take the form
\unskip\footnote{Here $\widehat{C}_2$ is the R-R potential for the part of $F_3$ that involves the isometries, while $\widetilde{F}_3 = \extder \widetilde{C}_2$ involves no isometries.
In principle, there is also a contribution to $F_5$ depending on $\widetilde{C}_{2}$.
However, we can always choose a gauge in which this contribution vanishes.}
\[\label{eq:rrfluxes5}
F_3 & = \extder C_2
& \qquad F_5 & = \extder C_4 + H \wedge C_2
\\ & = \extder \widehat{C}_2 + \widetilde{F}_3 ~,
& & = (1+\star) \big(\extder C_{4|t} + H \wedge \widehat{C}_{2|t} \big) ~,
\]
where $A_{n|t}$ denotes the part of the $n$-form $A_n$ that goes like $\extder t$, that is $A_{n|t} = \extder t \wedge \intder_\ind{T} A_n$ with $T^\mu = \delta^\mu_t$.
Writing
\[\label{eq:rrpotentials5}
& \begin{aligned}
\widehat{C}_2 = \frac{e^{-\Phi_0}}{2 \kappa \sqrt{1+\kappa^2} (N_+ N_-)^{\frac{1}{2}}} \frac{1}{2} c_{\mu_1 \mu_2} \, \extder X^{\mu_1} \wedge \extder X^{\mu_2} ~,
\end{aligned}
\\
& \begin{aligned}
\widetilde{F}_3 = \frac{2e^{-\Phi_0} \kappa^3 \sqrt{1+\kappa^2} }{(N_+ N_-)^{\frac{3}{2}}}
\frac{1}{3!} f_{\mu_1 \mu_2 \mu_3} \,\extder X^{\mu_1} \wedge \extder X^{\mu_2} \wedge \extder X^{\mu_3} ~,
\end{aligned}
\\
& \begin{aligned}
C_{4|t} & = \frac{e^{-\Phi_0}}{2 \sqrt{1+\kappa^2}(N_+ N_-)^{\frac{1}{2}}} \frac{1}{3!} c_{t \mu_1 \mu_2 \mu_3} \,\extder t \wedge \extder X^{\mu_1} \wedge \extder X^{\mu_2} \wedge \extder X^{\mu_3} ~,
\end{aligned}
\]
where $c_{\mu_1 \mu_2}$, $f_{\mu_1 \mu_2 \mu_3}$ and $c_{t\mu_1 \mu_2 \mu_3}$ are completely antisymmetric in the indices $\mu_i$, the non-vanishing components are
\begin{align}
c_{t\phi} & = + \Nn + 2\kappa^2 (1+\rho^2)(1-r^2) ~,
\nln
c_{\psi_1\phi_1} & = -\Nn + 2 (1-\kappa^2\rho^2)(1+\kappa^2 r^2) ~,
\nln
c_{\psi_2\phi_2} & = +\Nn - 2\kappa^2 \rho^2 r^2 x^2 w^2(1-\kappa^2\rho^2)(1+\kappa^2 r^2) ~,
\nln
c_{t\psi_2} & = -\Nn + 2\kappa^2 \rho^2 x^2 (1+\rho^2) (1+\kappa^2 r^2) ~,
\nln
c_{t\phi_2} & = +\Nn + 2\kappa^2 r^2 w^2 (1+\rho^2) (1+\kappa^2 r^2) ~,
\nln
c_{\psi_2 \phi} & = -\Nn + 2\kappa^2 \rho^2 x^2 (1-r^2) (1-\kappa^2 \rho^2) ~,
\nln
c_{\phi\phi_2} & = -\Nn - 2\kappa^2 r^2 w^2 (1-r^2) (1-\kappa^2 \rho^2) ~,
\nln & \nln[-4mm]
f_{\rho x r} & = + \rho^3 r x(1+\kappa^2r^4 w^2)\big(\Nn - 2 w^2(1-\kappa^2 \rho^2)(1+\kappa^2 r^2)\big) ~,
\nln
f_{\rho r w} & = + \rho r^3 w(1+\kappa^2\rho^4 x^2)\big(\Nn - 2 x^2(1-\kappa^2 \rho^2)(1+\kappa^2 r^2)\big) ~,
\nln
f_{\rho x w} & = + \rho^3 r^2 x w (1+\kappa^2 r^2)\big(r^2 \Nn - 2(1-r^2)(1-\kappa^2\rho^2)\big) ~,
\nln
f_{x r w} & = -\rho^2 r^3 x w (1-\kappa^2 \rho^2)\big(\rho^2 \Nn + 2(1+\rho^2)(1+\kappa^2r^2)\big) ~,
\nln & \nln[-4mm]
c_{t \psi_1 \psi_2 \rho} & = +\frac{\rho}{1-\kappa^2 \rho^2}\big(\Nn - 2 x^2 (1-\kappa^2 \rho^2)(1+\kappa^2 r^2)\big) ~,
\nln
c_{t \psi_1 \psi_2 x} & = - \frac{\rho^2 x}{1+\kappa^2 \rho^4 x^2}\big(\rho^2 \Nn + 2 (1+\rho^2)(1+\kappa^2 r^2)\big) ~,
\nln
c_{t \psi_2 \phi_1 r} & = +\frac{1}{\kappa^2 r(1-r^2)}\big(\Nn + 2 \kappa^4 \rho^2 r^2 x^2 (1+\rho^2)(1- r^2)\big) ~,
\nln
c_{t \psi_2 \phi_1 w} & = +\frac{1}{\kappa^2 w(1+\kappa^2 r^4 w^2)}\big(\Nn + 2 \kappa^4 \rho^2 r^4 x^2 w^2 (1+\rho^2)(1+\kappa^2 r^2)\big) ~,
\nln
c_{t \psi_1 \phi \rho} & = - \frac{\rho}{(1-\kappa^2 \rho^2 )(1-\kappa^2 \rho^2r^2 x^2 )} \big((1-r^2x^2)\Nn - 2(1-r^2)(1-\kappa^2\rho^2)(1+\kappa^2 r^2 x^2)\big) ~,
\nln
c_{t \psi_1 \phi x} & = + \frac{\rho^2 x (\rho^2 + r^2)}{(1+\kappa^2 \rho^4x^2)(1-\kappa^2 \rho^2 r^2 x^2 )} \big(\Nn + 2 \kappa^2 (1+\rho^2)(1-r^2)\big) ~,
\\
c_{t \psi_1 \phi r} & = + \frac{r}{(1+\kappa^2 r^2)(1-\kappa^2 \rho^2 r^2 x^2 )} \big((1+\rho^2 x^2)\Nn - 2 (1+\rho^2)(1+\kappa^2r^2)(1-\kappa^2 \rho^2 x^2)\big) ~,
\nln
c_{t \psi_1 \phi_2 \rho} & = - \frac{1+\kappa^2 r^2}{\kappa^2 \rho\big(1-\kappa^2( \rho^2-r^2-\rho^2 r^2 )\big)} \big(\Nn + 2 \kappa^4 \rho^2 r^2 w^2 (1+\rho^2)(1-r^2)\big) ~,
\nln
c_{t \psi_1 \phi_2 x} & = - \frac{1}{\kappa^2 x(1+\kappa^2 \rho^4 x^2)} \big(\Nn - 2 \kappa^4 \rho^4 r^2 x^2 w^2 (1+\rho^2)(1+\kappa^2 r^2)\big) ~,
\nln
c_{t \psi_1 \phi_2 r} & = + \frac{(1+\kappa^2)r}{\kappa^2 (1-r^2)\big(1-\kappa^2( \rho^2-r^2-\rho^2 r^2 )\big)} \big(\Nn + 2 \kappa^4 \rho^2 r^2 w^2 (1+\rho^2)(1-r^2)\big) ~,
\nln
c_{t \phi \phi_1 \rho} & = - \frac{\rho}{(1-\kappa^2 \rho^2)(1-\kappa^2 \rho^2 r^2 w^2)} \big((1-r^2 w^2)\Nn - 2 (1-r^2) (1-\kappa^2 \rho^2)(1+\kappa^2 r^2 w^2)\big) ~,
\nln
c_{t \phi \phi_1 r} & = + \frac{r}{(1+\kappa^2 r^2)(1-\kappa^2 \rho^2 r^2 w^2)} \big((1+\rho^2 w^2)\Nn - 2 (1+\rho^2) (1+\kappa^2 r^2)(1-\kappa^2 \rho^2 w^2)\big) ~,
\nln
c_{t \phi \phi_1 w} & = + \frac{r^2 w(\rho^2 + r^2)}{(1+\kappa^2 r^4 w^2)(1-\kappa^2 \rho^2 r^2 w^2)} \big(\Nn + 2 \kappa^2 (1+\rho^2)(1-r^2)\big) ~,
\nln
c_{t \phi_1 \phi_2 \rho} & = - \frac{(1+\kappa^2)\rho r^2 }{(1-\kappa^2 \rho^2)\big(1-\kappa^2(\rho^2-r^2-\rho^2r^2)\big)} \big(\Nn - 2w^2 (1-\kappa^2 \rho^2)(1+\kappa^2 r^2)\big) ~,
\nln
c_{t \phi_1 \phi_2 r} & = - \frac{r(1+ \rho^2)}{1-\kappa^2(\rho^2-r^2-\rho^2r^2)} \big(\Nn - 2w^2 (1-\kappa^2 \rho^2)(1+\kappa^2 r^2)\big) ~,
\nln
c_{t \phi_1 \phi_2 w} & = - \frac{r^2 w}{1+\kappa^2 r^4 w^2} \big(r^2\Nn - 2 (1+ \rho^2)(1+\kappa^2 r^2)\big) ~,
\nn
\end{align}
and
\[
N_\pm & = (1-\kappa^2\rho^2)(1+\kappa^2r^2)
\\ & \qquad + \kappa^2\rho^2r^2\big(\sqrt{1+\kappa^2}\sqrt{1-x^2}\sqrt{1-w^2}\pm\kappa\sqrt{1+\rho^2}\sqrt{1-r^2}x w\big)^2 ~,
\\
\Nn & = (1-\kappa^2\rho^2)(1+\kappa^2r^2)
\\ & \qquad + \kappa^2\rho^2r^2\big((1+\kappa^2)(1-x^2)(1-w^2) - \kappa^2 (1+\rho^2) (1-r^2) x^2 w^2\big) ~.
\]
As expected, this background satisfies the type IIB supergravity equations with the dilaton
\[\label{eq:dilaton5}
e^{-2\Phi} & = e^{-2\Phi_0} \frac{(1-\kappa^2 \rho^2)(1+\kappa^2 r^2)(1+ \kappa^2 \rho^4 x^2 )(1+\kappa^2 r^4 w^2)}{N_+ N_-} ~,
\]
and $\epsilon_{t\psi_1\psi_2\rho x \phi\phi_1\phi_2 r w} = - 1$.

The analytic continuations that correspond to considering the remaining two inequivalent R-matrices of $\mathfrak{su}(2,2)$, that is those associated with the two non-trivial permutations in equation \eqref{eq:su22permutations}, are \cite{Hoare:2016ibq}
\[\begin{gathered}\label{eq:analytic5}
t \to \psi_1 ~, \quad \psi_1 \to t ~, \quad \psi_2 \to \psi_2 ~, \quad \rho \to i \sqrt{1+\rho^2} ~, \quad x \to i x ~,
\\
t \to \psi_2 ~, \quad \psi_1 \to \psi_1 ~, \quad \psi_2 \to t ~, \quad \rho \to i \sqrt{1+\rho^2} ~, \quad x \to \sqrt{1+x^2} ~.
\end{gathered}\]
It is simple to check that both these analytic continuations preserve the reality of the background \eqref{eq:back5}, \eqref{eq:rrfluxes5}, \eqref{eq:dilaton5}.

\subsection{Limits}

To conclude this section we briefly discuss four limits of the supergravity background \eqref{eq:back5}, \eqref{eq:rrfluxes5}, \eqref{eq:dilaton5}.

In the $\kappa \to 0$ limit we expect to recover the maximally supersymmetric $\AdS_5 \times \Sp^5$ supergravity background.
This is not manifest in the expressions for the R-R fluxes given above.
For example, various components of the R-R potentials diverge in this limit.
However, this is an artifact of our choice of gauge and it is straightforward to check that the $\kappa \to 0$ limit of the R-R fluxes indeed gives the expected result.

The plane-wave limit \cite{Berenstein:2002jq,Blau:2002dy}, given by equations \eqref{eq:limpp2}, \eqref{eq:limpp1} and taking $L \to \infty$, results in the following supergravity background
\[
\extder s^2 &= -4 \extder x^- \extder x^+ - \mu^2 (1+\kappa^2) (\rho^2 + r^2) \, (\extder x^+)^2+ \extder \rho^2 + \rho^2 \extder \Omega_{3} + \extder r^2 + r^2 \extder \Omega_{3}'~, \\
\extder \Omega_{3} &=(1-x^2) \extder \psi_1^2 + \frac{\extder x^2}{1-x^2} + x^2 \extder \psi_2^2~, \qquad \extder \Omega_3' = (1-w^2) \extder \phi_1^2 + \frac{\extder w^2}{1-w^2} + w^2 \extder \phi_2^2~, \\
B &= \mu \kappa ( \rho \, \extder x^+ \wedge \extder \rho + r \, \extder x^+ \wedge \extder r)~, \qquad e^{-2 \Phi} = e^{-2 \Phi_0}~, \qquad F_3 =0~, \\
F_5 &= 4 \mu \sqrt{1+\kappa^2} ( \rho^3 x \, \extder x^+ \wedge \extder \psi_1 \wedge \extder \psi_2 \wedge \extder \rho \wedge \extder x -r^3 w \, \extder x^+ \wedge \extder \phi_1 \wedge \extder \phi_2 \wedge \extder r \wedge \extder w )~,
\]
which is identical to the plane-wave limit of the generalised supergravity background of \cite{Arutyunov:2015qva}, as discussed in \cite{Roychowdhury:2018qsz}.
Further taking $\mu \to 0$ gives flat space (in angular coordinates) with vanishing NS-NS and R-R fluxes.
As for the $\AdS_2 \times \Sp^2 \times \To^6$ case, we can reach the flat space background directly by rescaling $t \to L^{-1} t$ and $\phi \to L^{-1} \phi$ before taking $L \to \infty$, rather than using \eqref{eq:limpp2}.

The maximal deformation limit \cite{Arutyunov:2014cra,Pachol:2015mfa}, given by equation \eqref{eq:limmaxdef} and taking $\kappa \to \infty$, is finite, however, it does not match the mirror $\AdS_5 \times \Sp^5$ supergravity background of \cite{Arutyunov:2014cra,Arutyunov:2014jfa}.
It is also different to the maximal deformation limit of the generalised supergravity background of \cite{Arutyunov:2015qva}, which remains a generalised supergravity background.
Additionally rescaling the coordinates as in equation \eqref{eq:flatlimit} and taking $L \to \infty$ we recover flat space with vanishing NS-NS and R-R fluxes.
Indeed, the metric and B-field of the maximal deformation limit describe an integrable deformation of flat space with $\kappa$-deformed $\alg{iso}(1,4) \dsum \alg{iso}(5)$ symmetry \cite{Borowiec:2015wua,Pachol:2015mfa}.

The Pohlmeyer limit \cite{Hoare:2014pna}, given by equation \eqref{eq:limpohlmeyer} and taking $\epsilon \to 0$, is also finite, up to a divergent part of the B-field that is a closed two-form.
However, the resulting supergravity background has an imaginary B-field and R-R three-form.
It is also worth noting that, in contrast to the $\AdS_2 \times \Sp^2 \times \To^6$ case, the limit $\kappa \to i$, taken without rescaling the coordinates $t$ and $\phi$, is not finite.
This discrepancy might be explained by the fact that the $\kappa \to i$ limit of $\sdet(O_+)$ diverges in the $\AdS_5 \times \Sp^5$ case, but is finite in the $\AdS_2 \times \Sp^2 \times \To^6$ case.

\section{Discussion}\label{sec:discussion}

In this paper we have investigated the $\eta$-deformation of the $\AdS_2 \times \Sp^2 \times \To^6$ and $\AdS_5 \times \Sp^5$ superstrings for different Drinfel'd-Jimbo R-matrices.
In both cases we found that R-matrices associated with the Dynkin diagram that has all fermionic simple roots satisfy the unimodularity condition \eqref{eq:unimodularity}.
This is the requirement for the background to satisfy the type II supergravity equations \cite{Borsato:2016ose}.
It would be interesting to classify those Dynkin diagrams of basic Lie superalgebras that lead to R-matrices satisfying the unimodularity condition.

For $\AdS_2 \times \Sp^2 \times \To^6$ we considered the three Dynkin diagrams of $\alg{psl}(2|2;\Complex)$, explicitly showing that only when the simple roots are all fermionic do we find a supergravity background.
This background is a particular case of that found in \cite{Lunin:2014tsa}.
For $\AdS_5 \times \Sp^5$ we considered one particular Cartan-Weyl basis with all fermionic simple roots.
We constructed the corresponding R-R fluxes and dilaton that support the metric and B-field of $\eta$-deformed $\AdS_5 \times \Sp^5$ \cite{Arutyunov:2013ega} and confirmed that they satisfy the type IIB supergravity equations.

Let us emphasise that inequivalent Dynkin diagrams will typically lead to different R-R fluxes supporting the same metric and B-field within generalised supergravity.
For the two cases we have considered, demanding Weyl invariance, that is the background solves the standard type II supergravity equations, picks out the Dynkin diagram that has all fermionic simple roots as special.
However, it is worth noting that in the interesting recent work \cite{Fernandez-Melgarejo:2018wpg} progress has been made towards understanding the status of Weyl invariance for string sigma models on backgrounds solving the generalised type II supergravity equations, based on earlier results contained in \cite{Sakamoto:2017wor,Hoare:2015gda}.

In both the $\AdS_2 \times \Sp^2 \times \To^6$ and $\AdS_5 \times \Sp^5$ cases the maximal deformation limit of the supergravity background is finite, but does not recover the mirror supergravity background of \cite{Arutyunov:2014cra,Arutyunov:2014jfa}.
It therefore remains an open question to understand the connection between these models.
In the Pohlmeyer limit of the $\eta$-deformed $\AdS_2 \times \Sp^2 \times \To^6$ background we recovered the pp-wave supergravity background of \cite{Hoare:2014pna}, whose light-cone gauge-fixing gives the Pohlmeyer-reduced theory of the $\AdS_2 \times \Sp^2$ superstring \cite{Grigoriev:2007bu}.
The Pohlmeyer limit of the $\eta$-deformed $\AdS_5 \times \Sp^5$ background can also be taken, however in this case the resulting supergravity background is not real.
Understanding the relation between this background and the Pohlmeyer-reduced theory of the $\AdS_5 \times \Sp^5$ superstring \cite{Grigoriev:2007bu} also remains an open question.

\medskip

There are a number of AdS supergravity backgrounds that are described by the semi-symmetric space sigma model \cite{Zarembo:2010sg,Wulff:2014kja}.
It would be interesting to investigate $\eta$-deformations of these models for different R-matrices.
One example is the $\AdS_3 \times \Sp^3 \times \To^4$ superstring, which has superisometry algebra $\alg{psu}(1,1|2)\oplus\alg{psu}(1,1|2)$.
As this is two copies of the superisometry algebra of the $\AdS_2 \times \Sp^2 \times \To^6$ superstring it follows that R-matrices associated with the Dynkin diagram that has all fermionic simple roots will satisfy the unimodularity condition \eqref{eq:unimodularity}.
A candidate for a corresponding type IIB supergravity background is given in \cite{Lunin:2014tsa}.

When the superisometry algebra of the semi-symmetric space sigma model is of the form $\alg{g}\oplus \alg{g}$, a Wess-Zumino-Witten term can be added that corresponds to introducing NS-NS flux \cite{Cagnazzo:2012se}.
Such models also admit a bi-Yang-Baxter deformation \cite{Klimcik:2008eq,Klimcik:2014bta} with a different R-matrix and deformation parameter for each copy of $\alg{g}$ \cite{Hoare:2014oua}.
These can be combined into a three-parameter deformation of the semi-symmetric space sigma model \cite{Delduc:2018xug}.
It is reasonable to propose that when the two R-matrices satisfy the unimodularity condition \eqref{eq:unimodularity} this model is also Weyl invariant.
Subject to this being the case one could then study the corresponding supergravity backgrounds.
A candidate for the type IIB supergravity background corresponding to the bi-Yang-Baxter deformation of the $\AdS_3 \times \Sp^3 \times \To^4$ superstring is given in \cite{Lunin:2014tsa}.

\medskip

It was shown in \cite{Alvarez:1994np,Elitzur:1994ri} that non-abelian duality with respect to a bosonic Lie algebra leads to a Weyl anomaly when the trace of the structure constants is non-vanishing.
The anomaly is associated to integrating out the degrees of freedom of this non-unimodular algebra.
While the $\eta$-deformation is not equivalent to a non-abelian duality transformation, the results of this paper suggest that non-abelian duality with respect to a Lie superalgebra leads to a Weyl anomaly when the supertrace of the structure constants is non-vanishing.
It would be interesting to confirm this by direct computation.
See \cite{Borsato:2018idb} for recent progress in this direction.

In order to better understand the implications of considering inequivalent Dynkin diagrams it would be useful to investigate the Poisson-Lie symmetry \cite{Klimcik:1995ux,Klimcik:1995jn} and the associated $q$-deformed superisometry algebra \cite{Delduc:2014kha,Delduc:2013fga,Delduc:2016ihq} in more detail.
Models with Poisson-Lie symmetry can be dualised with respect to this symmetry.
A systematic way of performing this duality is to start from a first-order action on the Drinfel'd double given by the complexified superisometry algebra and integrate out the degrees of freedom of different maximally isotropic subalgebras \cite{Klimcik:1995dy,Klimcik:1996nq}.
Indeed, generalising the results of non-abelian duality, there is evidence that integrating out the degrees of freedom of a non-unimodular algebra is also associated to a Weyl anomaly in models with Poisson-Lie symmetry \cite{Tyurin:1995bu,Bossard:2001au}.
With this in mind, it may prove insightful to investigate the relation between different possible Poisson-Lie duals of the $\eta$-deformed semi-symmetric space sigma model for R-matrices associated with inequivalent Dynkin diagrams in the spirit of \cite{Hoare:2018ebg}.

\medskip

A conjecture for the light-cone gauge-fixed S-matrix of the $\eta$-deformed $\AdS_5 \times \Sp^5$ superstring based on symmetries has been given in \cite{Beisert:2008tw,Hoare:2011wr} and the resulting finite-size spectrum analysed in \cite{Arutynov:2014ota,Klabbers:2017vtw}.
However, using the reference R-matrix \eqref{eq:refrmatrix} associated with the distinguished Dynkin diagram, the perturbative computation does not match the expansion of the exact result \cite{Arutyunov:2015qva}.
It is therefore natural to ask whether instead using an R-matrix associated with the Dynkin diagram that has all fermionic simple roots provides a resolution to this disagreement.
Finally, understanding the singularity of the backgrounds at $\rho = \kappa^{-1}$ (or for the two backgrounds given by the analytic continuations \eqref{eq:analytic5} at $x = \kappa^{-1} (1+\rho^2)^{-1}$ and $\rho \to \infty$ respectively) remains an important open problem.

\section*{Acknowledgements}

We would like to thank L.~Wulff for helpful discussions and R.~Borsato, M.~Magro, R.~Roiban, A.~A.~Tseytlin, S.~van~Tongeren, B.~Vicedo and L.~Wulff for comments on the draft.
This work is supported by grant no.~615203 from the European Research Council under the FP7.

\appendix

\section{Conventions for \texorpdfstring{$\AdS_2 \times \Sp^2 \times \To^6$}{AdS2 x S2 x T6}}\label{app:ads2}

In this appendix we give our conventions for the gamma matrices, superalgebras and R-matrices used in the construction of the (generalised) supergravity backgrounds corresponding to the $\eta$-deformation of the $\AdS_2 \times \Sp^2 \times \To^6$ superstring.

\subsection{Gamma matrices}\label{app:gamma}

\paragraph{$4$-dimensional gamma matrices.}
Starting from the Pauli matrices
\[\begin{gathered}
\sigma_1 = \begin{pmatrix} 0 & 1 \\ 1 & 0 \end{pmatrix}~, \quad \sigma_2 = \begin{pmatrix} 0 & -i \\ i & 0 \end{pmatrix}~, \quad \sigma_3 = \begin{pmatrix} 1 & 0 \\ 0 & -1 \end{pmatrix}~,
\qquad \sigma_\pm = \frac12(\sigma_1 \pm i \sigma_2) ~,
\end{gathered}\]
we define the $4$-dimensional gamma matrices
\[
\bar{\gamma}^0 &= -i \sigma_3 \otimes \identity_2~, &\qquad \bar{\gamma}^1 &= \sigma_1 \otimes \identity_2~, \qquad
\bar{\gamma}^2 &= -\identity_2 \otimes i\sigma_3~, &\qquad \bar{\gamma}^3 &= -\identity_2 \otimes i\sigma_1~.
\]
These gamma matrices do not satisfy the Clifford algebra in $1+3$ dimensions, however $\set{\bar{\gamma}^0,\bar{\gamma}^1}$ satisfy the Clifford algebra in $1+1$ dimensions and $\set{\bar{\gamma}^2,\bar{\gamma}^3}$ in $2$ dimensions.

\paragraph{$32$-dimensional gamma matrices.}
We choose the following representation for the ten $32 \times 32$ gamma matrices appearing in the Green Schwarz action
\[
\Gamma^0 &= -i \sigma_1 \otimes \identity_2 \otimes \identity_2 \otimes \sigma_3 \otimes \identity_2~, & \qquad
\Gamma^1 &= \sigma_1 \otimes \identity_2 \otimes \identity_2 \otimes \sigma_1 \otimes \identity_2~, \\
\Gamma^2 &= \sigma_2 \otimes \identity_2 \otimes \identity_2 \otimes \identity_2 \otimes \sigma_3~, & \qquad
\Gamma^3 &= \sigma_2 \otimes \identity_2 \otimes \identity_2 \otimes \identity_2 \otimes \sigma_1~, \\
\Gamma^4 &= \sigma_1 \otimes \sigma_1 \otimes \sigma_2 \otimes \sigma_2 \otimes \identity_2~, & \qquad
\Gamma^5 &= -i\sigma_2 \otimes \sigma_2 \otimes \sigma_1 \otimes \identity_2 \otimes \sigma_2~, \\
\Gamma^6 &= \sigma_1 \otimes U^{-1} (\sigma_2 \otimes \identity_2) U \otimes \sigma_2 \otimes \identity_2~, & \qquad
\Gamma^7 &= -i\sigma_2 \otimes U^{-1} (\identity_2 \otimes \sigma_2) U \otimes \identity_2 \otimes \sigma_2~, \\
\Gamma^8 &= \sigma_1 \otimes U^{-1}(\sigma_3 \otimes \sigma_2)U \otimes \sigma_2 \otimes \identity_2~, & \qquad
\Gamma^9 &= -i\sigma_2 \otimes U^{-1}(\sigma_2 \otimes \sigma_3) U \otimes \identity_2 \otimes \sigma_2 ~,
\]
with
\[
U= \frac{1}{\sqrt{2}} \begin{pmatrix}
\sqrt{2} & 0 & 0 & 0 \\
0 & 1 & 1 & 0 \\
0 & -1 & 1 & 0 \\
0 & 0 & 0 & \sqrt{2}
\end{pmatrix}~.
\]
They satisfy the Clifford algebra in $1+9$ dimensions and are related to the $4$-dimensional gamma matrices by
\[
\Gamma^a = \sigma_1 \otimes \identity_4 \otimes \bar{\gamma}^a ~, \quad a=0,1 ~,
\qquad \Gamma^a = -i \sigma_2 \otimes \identity_4 \otimes \bar{\gamma}^a~,\quad a=2,3~.
\]
Furthermore, we have
\[
\Gamma^{11} = \Gamma^{0} \Gamma^{1} \Gamma^{2} \Gamma^{3} \Gamma^{4} \Gamma^{5} \Gamma^{6} \Gamma^{7} \Gamma^{8} \Gamma^{9} = \sigma_3 \otimes \identity_{16}~.
\]

\paragraph{$16$-dimensional chiral gamma matrices.}
The $16 \times 16$ chiral gamma matrices are then identified through
\[ \Gamma^a =
\begin{pmatrix}
0 & (\gamma^a)^{\alpha \beta} \\
(\gamma^a)_{\alpha \beta} & 0
\end{pmatrix} ~,
\]
and satisfy $\gamma^a_{\alpha \beta} (\gamma^b)^{\beta \gamma} + \gamma^b_{\alpha \beta} (\gamma^a)^{\beta \gamma} =2 \eta^{ab} \delta_\alpha^\gamma $.
The projector
\[
\mathcal{P}_4 = \frac{1}{4} (\identity_{16} - \gamma^{4567} - \gamma^{4589} - \gamma^{6789}) = \diag(0,1,0,0) \otimes \identity_4 ~,
\]
appearing in the R-R bispinor \eqref{eq:bispinorAdS2} projects onto a $4$-dimensional spinor subspace and thus can be used to effectively make these matrices $4 \times 4$ with spinor index $\alpha=1,2,3,4$.
In particular, we have
\[
\mathcal{P}_4 \gamma^a \mathcal{P}_4 &\rightarrow \bar{\gamma}^a~, & \quad a & =0,1,2,3~, \\
\mathcal{P}_4 \gamma^a \mathcal{P}_4 &\rightarrow 0~, & \quad a & = 4,\dots,9~,
\]
where the arrow represents the projection onto $4$ dimensions.

\subsection{The complexified superalgebra \texorpdfstring{$\alg{sl}(2|2;\Complex)$}{sl(2|2;C)}}

The bosonic subalgebra of $\alg{sl}(2|2;\Complex)$ is $\alg{sl}(2;\Complex) \dsum \alg{sl}(2;\Complex) \dsum \alg{gl}(1;\Complex)$.
We introduce the corresponding generators $\gen{K}_0, \gen{K}_\pm$, $\gen{L}_0, \gen{L}_\pm$ and $\gen{C}_0$ along with the eight supercharges $\gen{Q}^{\pm\check{\ind{A}} \hat{\ind{A}}}$.
Here $\check{A} = \pm$ is the spinor index associated with the first copy of $\alg{sl}(2;\Complex)$ and $\hat{A} = \pm$ to the second.
The first index on the supercharges corresponds to their splitting under the $\alg{gl}(1;\Complex)$ outer automorphism
\[
\com{\gen{R}}{\gen{Q}^{\pm\check{\ind{A}} \hat{\ind{A}}}} = \pm \half \gen{Q}^{\pm\check{\ind{A}} \hat{\ind{A}}}~.
\]
The non-vanishing commutation and anti-commutation relations are
\[
& \com{\gen{K}_0}{\gen{K}_\pm} = \pm \gen{K}_\pm~, && \com{\gen{K}_+}{\gen{K}_-} = 2\gen{K}_0~,
\\
&\com{\gen{L}_0}{\gen{L}_\pm} = \pm \gen{L}_\pm~, && \com{\gen{L}_+}{\gen{L}_-} = 2 \gen{L}_0~,
\\
& \com{\gen{K}_0}{\gen{Q}^{\ind{B}\pm \hat{\ind{A}}}} = \pm \half \gen{Q}^{\ind{B}\pm \hat{\ind{A}}}~,&& \com{\gen{K}_\pm}{\gen{Q}^{\ind{B}\mp \hat{\ind{A}}}} = \gen{Q}^{\ind{B}\pm \hat{\ind{A}}}~,
\\
&\com{\gen{L}_0}{\gen{Q}^{\ind{B} \check{\ind{A}} \pm }} = \pm \half \gen{Q}^{\ind{B} \check{\ind{A}} \pm }~,&& \com{\gen{L}_\pm}{\gen{Q}^{ \ind{B}\check{\ind{A}} \mp }} = \gen{Q}^{\ind{B}\check{\ind{A}} \pm }~,
\\
&\anticom{\gen{Q}^{+ \pm +}}{\gen{Q}^{-\pm -}} = \pm \gen{K}_\pm~, && \anticom{\gen{Q}^{- \pm +}}{\gen{Q}^{+\pm -}} = \mp \gen{K}_\pm~, \\
& \anticom{\gen{Q}^{+ + \pm}}{\gen{Q}^{- - \pm}} = \mp \gen{L}_\pm~, && \anticom{\gen{Q}^{- + \pm}}{\gen{Q}^{+ - \pm}} = \pm \gen{L}_\pm~,
\\
& \anticom{\gen{Q}^{\pm+ \pm}}{\gen{Q}^{\mp- \mp}} = - \gen{K}_0 \pm \gen{L}_0 \mp \gen{C}_0~, \qquad
&& \anticom{\gen{Q}^{\mp + \pm}}{\gen{Q}^{\pm - \mp}} = + \gen{K}_0 \mp \gen{L}_0 \mp \gen{C}_0~.
\]
The central element $\gen{C}_0$ commutes with all generators.

\paragraph{Cartan-Weyl basis.}
The three Dynkin diagrams of $\alg{sl}(2;\Complex)$ correspond to inequivalent sets of simple roots.
To identify the roots, let us introduce a generic Cartan-Weyl basis for $\alg{sl}(2|2 ; \Complex)$ composed of the three Cartan generators $\set{h_i}$ and the positive $\set{e_i}$ and negative $\set{f_i}$ simple roots satisfying the defining relations
\[
\com{h_i}{e_j} = a_{ij} e_j ~, \qquad \com{h_i}{f_j} = - a_{ij} f_j ~, \qquad [e_i,f_j\} = \delta_{ij} h_j ~,
\]
where $a_{ij}$ is the symmetrised Cartan matrix.
The non-simple roots are
\[
e_{12} &= [e_1,e_2\}~, &\qquad e_{23} &= [e_2,e_3\}~, &\qquad e_{123} &= [e_1,[e_2,e_3\}\}~, \\
f_{21} &= [f_2,f_1\}~, &\qquad f_{32} &= [f_3,f_2\}~, &\qquad f_{321} &= [f_3,[f_2,f_1\}\}~.
\]

\paragraph{Matrix Realisation.}
To define the action of the R-matrix we use the following matrix realisation of the complexified superalgebra $\alg{sl}(2|2;\Complex)$
\[
& \gen{K}_0 = -\frac{1}{2} \begin{pmatrix}\sigma_3 & 0 \\ 0 & 0 \end{pmatrix} ~,
&& \gen{K}_\pm = \begin{pmatrix}\sigma_\mp & 0 \\ 0 & 0 \end{pmatrix} ~,
\\
& \gen{L}_0 = \frac{1}{2} \begin{pmatrix}0 & 0 \\ 0 & \sigma_3 \end{pmatrix} ~,
&& \gen{L}_\pm = \begin{pmatrix}0 & 0 \\ 0 & \sigma_\pm \end{pmatrix}~.
\\
& \gen{Q}^{+++} = \begin{pmatrix} 0 & -N_{22} \\ 0 & 0 \end{pmatrix}~, \qquad \qquad
&& \gen{Q}^{---} = \begin{pmatrix} 0 & 0 \\ N_{22} & 0 \end{pmatrix}~,
\\
& \gen{Q}^{++-} = \begin{pmatrix} 0 & N_{21} \\ 0 & 0 \end{pmatrix}~,
&& \gen{Q}^{--+} = \begin{pmatrix} 0 & 0 \\ N_{12} & 0 \end{pmatrix}~,
\\
& \gen{Q}^{+-+} = \begin{pmatrix} 0 & -N_{12} \\ 0 & 0 \end{pmatrix}~,
&& \gen{Q}^{-+-} = \begin{pmatrix} 0 & 0 \\ -N_{21} & 0 \end{pmatrix}~,
\\
& \gen{Q}^{+--} = \begin{pmatrix} 0 & N_{11} \\ 0 & 0 \end{pmatrix}~,
&& \gen{Q}^{-++} = \begin{pmatrix} 0 & 0 \\ -N_{11} & 0 \end{pmatrix}~,
\]
where
\[
(N_\ind{\check{\alpha} \hat{\alpha}})_\ind{\check{\beta} \hat{\beta}} = \delta_\ind{\check{\alpha} \check{\beta}} \delta_\ind{\hat{\alpha} \hat{\beta}}~, \qquad \check{\alpha},\check{\beta},\hat{\alpha},\hat{\beta}=1,2 ~.
\]
In particular, the generators $\gen{K}_-$, $\gen{L}_+$ and $\gen{Q}^{+ \check{\ind{A}} \hat{\ind{A}}}$ are upper-triangular matrices, while $\gen{K}_+$, $\gen{L}_-$ and $\gen{Q}^{- \check{\ind{A}} \hat{\ind{A}}}$ are lower-triangular.

\paragraph{Inequivalent R-matrices and associated Dynkin diagrams.}
The first Dynkin diagram we consider is $\Circle-\otimes-\Circle$ with two bosonic simple roots and one fermionic.
A choice of Cartan generators and positive and negative simple roots is
\[
&h_1 = 2 \gen{K}_0~, &\qquad & e_1 = - \gen{K}_-~, &\qquad & f_1 = \gen{K}_+~, \\
&h_2 = - \gen{K}_0 - \gen{L}_0 - \gen{C}_0 ~, &\qquad & e_2 = \gen{Q}^{++-}~, &\qquad & f_2 = - \gen{Q}^{--+} ~, \\
&h_3 = 2 \gen{L}_0~, &\qquad & e_3 = \gen{L}_+~, & \qquad & f_3 = \gen{L}_- ~.
\]
The associated R-matrix is $R_0$ \eqref{eq:r0}, the action of which is given in equation \eqref{eq:refrmatrix}.

The second Dynkin diagram we consider is $\otimes-\Circle-\otimes$ with one bosonic simple root and two fermionic.
A choice of the Cartan generators and positive and negative simple roots is
\[
&h_1 = \gen{K}_0 - \gen{L}_0 - \gen{C}_0 ~, &\qquad & e_1 = \gen{Q}^{+--} ~, &\qquad & f_1 = \gen{Q}^{-++} ~, \\
&h_2 = 2\gen{L}_0~, &\qquad & e_2 = \gen{L}_{+} ~, &\qquad & f_2 = \gen{L}_- ~, \\
&h_3 = \gen{K}_0 - \gen{L}_0 + \gen{C}_0 ~, &\qquad & e_3 = \gen{Q}^{---} ~, &\qquad & f_3 = -\gen{Q}^{+++} ~.
\]
The associated R-matrix is $R_1$ \eqref{eq:r1}, the action of which is the same as in equation \eqref{eq:refrmatrix} except with
\[
\epsilon = \begin{pmatrix}
0 & +1 & +1 & +1 \\
-1 & 0 & -1 & -1 \\
-1 & +1 & 0 & +1 \\
-1 & +1 & -1 & 0
\end{pmatrix}~.
\]

The third and final Dynkin diagram we consider is $\otimes-\otimes-\otimes$ with three fermionic simple roots.
A choice of the Cartan generators and positive and negative simple roots is
\[
&h_1 = -\gen{K}_0 + \gen{L}_0 - \gen{C}_0 ~, &\qquad & e_1 = \gen{Q}^{+++} ~, &\qquad & f_1 = \gen{Q}^{---} ~, \\
&h_2 = \gen{K}_0 + \gen{L}_0 + \gen{C}_0~, &\qquad & e_2 = \gen{Q}^{--+} ~, &\qquad & f_2 = \gen{Q}^{++-} ~, \\
&h_3 = \gen{K}_0 - \gen{L}_0 - \gen{C}_0 ~, &\qquad & e_3 = \gen{Q}^{+--} ~, &\qquad & f_3 = \gen{Q}^{-++} ~.
\]
The associated R-matrix is $R_2$ \eqref{eq:r2}, the action of which is the same as in equation \eqref{eq:refrmatrix} except with
\[
\epsilon = \begin{pmatrix}
0 & +1 & +1 & +1 \\
-1 & 0 & -1 & +1 \\
-1 & +1 & 0 & +1 \\
-1 & -1 & -1 & 0
\end{pmatrix} ~.
\]

\subsection{The real form \texorpdfstring{$\alg{psu}(1,1|2)$}{psu(1,1|2)}}

The real form $\alg{su}(1,1|2)$ is given by those elements of the complexified superalgebra $\alg{sl}(2|2;\Complex)$ satisfying
\[ \label{eq:realitycond} M^\dagger H + H M =0~, \qquad H=\begin{pmatrix} \sigma_3 & 0 \\ 0 & \identity_2 \end{pmatrix}~.
\]
The superalgebra $\alg{su}(1,1|2)$ contains the $1$-dimensional ideal $\alg{u}(1)$ generated by $i \identity_4$.
The quotient of $\alg{su}(1,1|2)$ over this $\alg{u}(1)$ subalgebra defines the superalgebra $\alg{psu}(1,1|2)$.

The automorphism
\[\begin{gathered}
\Omega(M) = - K^{-1} M^{\supertranspose} K ~, \qquad K = \begin{pmatrix} \sigma_3 & 0 \\ 0 & \sigma_3 \end{pmatrix} ~,
\\
M^{\supertranspose}
= P_{\alg{su}(1,1)} M^t P_{\alg{su}(1,1)}
- P_{\alg{su}(1,1)} M^t P_{\alg{su}(2)}
+ P_{\alg{su}(2)} M^t P_{\alg{su}(1,1)}
+ P_{\alg{su}(2)} M^t P_{\alg{su}(2)} ~,
\end{gathered}\]
where
\[
P_{\alg{su}(1,1)} = \begin{pmatrix} \identity_2 & 0 \\ 0 & 0 \end{pmatrix}~, \qquad
P_{\alg{su}(2)} = \begin{pmatrix} 0 & 0 \\ 0 & \identity_2 \end{pmatrix}~,
\]
endows the $\alg{psu}(1,1|2)$ superalgebra with a $\Integer_4$ grading and the elements of grade $k$ satisfy $\Omega(M)=i^k M$.
The generators below are chosen so that they belong to a specific grading.

\paragraph{Bosonic generators.}
Our choice for the three $\alg{su}(1,1)$ generators is
\[
P_0 &= \frac{1}{2} \begin{pmatrix} i \sigma_3 & 0 \\ 0 & 0 \end{pmatrix}=-i \gen{K}_0 ~, \qquad
P_1 = \frac{1}{2} \begin{pmatrix} \sigma_2 & 0 \\ 0 & 0 \end{pmatrix} = \frac{i}{2} (\gen{K}_+-\gen{K}_-) ~,
\\
J_{01} &= - \com{P_0}{P_1}= -\frac{1}{2} \begin{pmatrix} \sigma_1 & 0 \\ 0 & 0 \end{pmatrix}= -\frac{1}{2}(\gen{K}_+ + \gen{K}_-)~, \\
\]
and for the three $\alg{su}(2)$ generators is
\[
P_2 &= \frac{1}{2} \begin{pmatrix} 0 & 0 \\ 0 & i \sigma_3 \end{pmatrix}= i \gen{L}_0 ~, \qquad
P_3 = \frac{1}{2} \begin{pmatrix} 0 & 0 \\ 0 & i \sigma_2 \end{pmatrix} = \frac{1}{2} (\gen{L}_+ - \gen{L}_-) ~,
\\
J_{23} &= + \com{P_2}{P_3}= \frac{1}{2} \begin{pmatrix} 0 & 0 \\ 0 & i \sigma_1 \end{pmatrix} = \frac{i}{2} (\gen{L}_+ + \gen{L}_-)~.
\]
Here $J_{01}$ and $J_{23}$ generate the $\alg{so}(1,1) \dsum \alg{so}(2)$ grade 0 subalgebra, while the other bosonic generators $P_a$, $a=0,1,2,3$ are of grade 2.

\paragraph{Fermionic generators.}
The $\alg{psu}(1,1|2)$ superalgebra also contains eight fermionic generators $Q_{\ind{I} \check{\alpha} \hat{\alpha}}$, where $I=1$ for generators of grade 1 and $I=2$ for grade 3, $\check{\alpha}=1,2$ is the spinor $\alg{su}(1,1)$ index and $\hat{\alpha}=1,2$ is the $\alg{su}(2)$ spinor index.
Explicitly, these generators are
\[
Q_{1 \check{\alpha} \hat{\alpha}} = \frac{1}{\sqrt{2}} i^{(\check{\alpha}-\hat{\alpha})} \begin{pmatrix} 0 & N_{\check{\alpha} \hat{\alpha}} \\ i \sigma_3 (N_{\check{\alpha} \hat{\alpha}})^t \sigma_3 & 0 \end{pmatrix}~,
\\
Q_{2 \check{\alpha} \hat{\alpha}} = \frac{1}{\sqrt{2}} i^{(\check{\alpha}-\hat{\alpha})} \begin{pmatrix} 0 & i N_{\check{\alpha} \hat{\alpha}} \\ \sigma_3 (N_{\check{\alpha} \hat{\alpha}})^t \sigma_3 & 0 \end{pmatrix}~.
\]
To make the link with the notation used in the main text and \appref{app:gamma}, the spinor indices $\check{\alpha}$ and $\hat{\alpha}$ can be gathered into a single index, $\alpha=1,2,3,4$, and we define the generators $Q_{\ind{I} \alpha}$ as
\[
Q_{11}= Q_{111}~, \qquad Q_{12}= Q_{112}~, \qquad Q_{13}= Q_{121}~, \qquad Q_{14}= Q_{122}~.
\]
While these generators themselves do not satisfy the reality condition \eqref{eq:realitycond}, elements of the Grassmann envelope $\theta^{I \alpha} Q_{I \alpha}$ will do so if one imposes suitable reality conditions on the 4-component fermions $\theta^{I \alpha}$.
Our choice of generators matches the conventions of \cite{Borsato:2016ose}, in particular we have the anti-commutation relations
\[
\anticom{Q_{1 \alpha}}{Q_{1 \beta}}= \anticom{Q_{2 \alpha}}{Q_{2 \beta}} = i \bar{\gamma}^a_{\alpha \beta} P_a~.
\]
\begin{bibtex}[\jobname]

@article{Metsaev:1998it,
author         = "Metsaev, R. R. and Tseytlin, Arkady A.",
title          = "{Type IIB superstring action in $AdS_5 \times S^5$ background}",
journal        = "Nucl. Phys.",
volume         = "B533",
year           = "1998",
pages          = "109-126",
doi            = "10.1016/S0550-3213(98)00570-7",
eprint         = "hep-th/9805028",
archivePrefix  = "arXiv",
primaryClass   = "hep-th",
reportNumber   = "FIAN-TD-98-21, IMPERIAL-TP-97-98-44, NSF-ITP-98-055",
SLACcitation   = "
}

@article{Berkovits:1999zq,
author         = "Berkovits, N. and Bershadsky, M. and Hauer, T. and Zhukov, S. and Zwiebach, B.",
title          = "{Superstring theory on $AdS_2 \times S^2$ as a coset supermanifold}",
journal        = "Nucl. Phys.",
volume         = "B567",
year           = "2000",
pages          = "61-86",
doi            = "10.1016/S0550-3213(99)00683-5",
eprint         = "hep-th/9907200",
archivePrefix  = "arXiv",
primaryClass   = "hep-th",
reportNumber   = "IFT-P-060-99, HUTP-99-A044, MIT-CTP-2878, CTP-MIT-2878",
SLACcitation   = "
}

@article{Borsato:2016ose,
author         = "Borsato, Riccardo and Wulff, Linus",
title          = "{Target space supergeometry of $\eta$ and $\lambda$-deformed strings}",
journal        = "JHEP",
volume         = "10",
year           = "2016",
pages          = "045",
doi            = "10.1007/JHEP10(2016)045",
eprint         = "1608.03570",
archivePrefix  = "arXiv",
primaryClass   = "hep-th",
reportNumber   = "IMPERIAL-TP-LW-2016-03",
SLACcitation   = "
}

@article{Arutyunov:2015mqj,
author         = "Arutyunov, G. and Frolov, S. and Hoare, B. and Roiban, R. and Tseytlin, A. A.",
title          = "{Scale invariance of the $\eta$-deformed $AdS_5 \times S^5$ superstring, T-duality and modified type II equations}",
journal        = "Nucl. Phys.",
volume         = "B903",
year           = "2016",
pages          = "262-303",
doi            = "10.1016/j.nuclphysb.2015.12.012",
eprint         = "1511.05795",
archivePrefix  = "arXiv",
primaryClass   = "hep-th",
reportNumber   = "ZMP-HH-15-27, TCDMATH-15-12, IMPERIAL-TP-AT-2015-08",
SLACcitation   = "
}

@article{Wulff:2016tju,
author         = "Tseytlin, A. A. and Wulff, L.",
title          = "{Kappa-symmetry of superstring sigma model and generalized 10d supergravity equations}",
journal        = "JHEP",
volume         = "06",
year           = "2016",
pages          = "174",
doi            = "10.1007/JHEP06(2016)174",
eprint         = "1605.04884",
archivePrefix  = "arXiv",
primaryClass   = "hep-th",
reportNumber   = "IMPERIAL-TP-LW-2016-02",
SLACcitation   = "
}

@article{Arutyunov:2013ega,
author         = "Arutyunov, Gleb and Borsato, Riccardo and Frolov, Sergey",
title          = "{S-matrix for strings on $\eta$-deformed $AdS_5 \times S^5$}",
journal        = "JHEP",
volume         = "04",
year           = "2014",
pages          = "002",
doi            = "10.1007/JHEP04(2014)002",
eprint         = "1312.3542",
archivePrefix  = "arXiv",
primaryClass   = "hep-th",
reportNumber   = "ITP-UU-13-31, SPIN-13-23, HU-MATHEMATIK-2013-24, TCD-MATH-13-16",
SLACcitation   = "
}

@article{Arutyunov:2015qva,
author         = "Arutyunov, Gleb and Borsato, Riccardo and Frolov, Sergey",
title          = "{Puzzles of $\eta$-deformed $AdS_5 \times S^5$}",
journal        = "JHEP",
volume         = "12",
year           = "2015",
pages          = "049",
doi            = "10.1007/JHEP12(2015)049",
eprint         = "1507.04239",
archivePrefix  = "arXiv",
primaryClass   = "hep-th",
reportNumber   = "ITP-UU-15-10, TCD-MATH-15-05, ZMP-HH-15-19",
SLACcitation   = "
}

@article{Arutynov:2014ota,
author         = "Arutyunov, Gleb and de Leeuw, Marius and van Tongeren, Stijn J.",
title          = "{The exact spectrum and mirror duality of the $(AdS_5 \times S^5)_\eta$ superstring}",
journal        = "Theor. Math. Phys.",
volume         = "182",
year           = "2015",
number         = "1",
pages          = "23-51",
doi            = "10.1007/s11232-015-0243-9",
note           = "[\textsf{Teor. Mat. Fiz. 182, 28 (2014)}]",
eprint         = "1403.6104",
archivePrefix  = "arXiv",
primaryClass   = "hep-th",
SLACcitation   = "
}

@article{Klabbers:2017vtw,
author         = "Klabbers, Rob and van Tongeren, Stijn J.",
title          = "{Quantum Spectral Curve for the $\eta$-deformed $AdS_5 \times S^5$ superstring}",
journal        = "Nucl. Phys.",
volume         = "B925",
year           = "2017",
pages          = "252-318",
doi            = "10.1016/j.nuclphysb.2017.10.005",
eprint         = "1708.02894",
archivePrefix  = "arXiv",
primaryClass   = "hep-th",
reportNumber   = "ZMP-HH-17-25, HU-EP-17-21",
SLACcitation   = "
}

@article{Arutyunov:2014cra,
author         = "Arutyunov, Gleb and van Tongeren, Stijn J.",
title          = "{$AdS_5 \times S^5$ mirror model as a string sigma model}",
journal        = "Phys. Rev. Lett.",
volume         = "113",
year           = "2014",
pages          = "261605",
doi            = "10.1103/PhysRevLett.113.261605",
eprint         = "1406.2304",
archivePrefix  = "arXiv",
primaryClass   = "hep-th",
reportNumber   = "HU-EP-14-21, HU-MATH-14-12, ITP-UU-14-18, SPIN-14-16",
SLACcitation   = "
}

@article{Arutyunov:2014jfa,
author         = "Arutyunov, Gleb and van Tongeren, Stijn J.",
title          = "{Double Wick rotating Green-Schwarz strings}",
journal        = "JHEP",
volume         = "05",
year           = "2015",
pages          = "027",
doi            = "10.1007/JHEP05(2015)027",
eprint         = "1412.5137",
archivePrefix  = "arXiv",
primaryClass   = "hep-th",
SLACcitation   = "
}

@article{Pachol:2015mfa,
author         = "Pacho\l{}, Anna and van Tongeren, Stijn J.",
title          = "{Quantum deformations of the flat space superstring}",
journal        = "Phys. Rev.",
volume         = "D93",
year           = "2016",
pages          = "026008",
doi            = "10.1103/PhysRevD.93.026008",
eprint         = "1510.02389",
archivePrefix  = "arXiv",
primaryClass   = "hep-th",
reportNumber   = "HU-EP-15-48, HU-MATH-15-13",
SLACcitation   = "
}

@article{Beisert:2008tw,
author         = "Beisert, Niklas and Koroteev, Peter",
title          = "{Quantum deformations of the one-dimensional Hubbard model}",
journal        = "J. Phys.",
volume         = "A41",
year           = "2008",
pages          = "255204",
doi            = "10.1088/1751-8113/41/25/255204",
eprint         = "0802.0777",
archivePrefix  = "arXiv",
primaryClass   = "hep-th",
reportNumber   = "AEI-2008-003, ITEP-TH-06-08",
SLACcitation   = "
}

@article{Hoare:2011wr,
author         = "Hoare, Ben and Hollowood, Timothy J. and Miramontes, J. Luis",
title          = "{q-deformation of the $AdS_5 \times S^5$ superstring S-matrix and its relativistic limit}",
journal        = "JHEP",
volume         = "03",
year           = "2012",
pages          = "015",
doi            = "10.1007/JHEP03(2012)015",
eprint         = "1112.4485",
archivePrefix  = "arXiv",
primaryClass   = "hep-th",
reportNumber   = "IMPERIAL-TP-11-BH-03",
SLACcitation   = "
}

@article{Hoare:2014pna,
author         = "Hoare, B. and Roiban, R. and Tseytlin, A. A.",
title          = "{On deformations of $AdS_n \times S^n$ supercosets}",
journal        = "JHEP",
volume         = "06",
year           = "2014",
pages          = "002",
doi            = "10.1007/JHEP06(2014)002",
eprint         = "1403.5517",
archivePrefix  = "arXiv",
primaryClass   = "hep-th",
reportNumber   = "IMPERIAL-TP-AT-2014-02, HU-EP-14-10",
SLACcitation   = "
}

@article{Lunin:2014tsa,
author         = "Lunin, O. and Roiban, R. and Tseytlin, A. A.",
title          = "{Supergravity backgrounds for deformations of $AdS_n \times S^n$ supercoset string models}",
journal        = "Nucl. Phys.",
volume         = "B891",
year           = "2015",
pages          = "106-127",
doi            = "10.1016/j.nuclphysb.2014.12.006",
eprint         = "1411.1066",
archivePrefix  = "arXiv",
primaryClass   = "hep-th",
reportNumber   = "IMPERIAL-TP-AT-2014-07",
SLACcitation   = "
}

@article{Delduc:2013qra,
author         = "Delduc, Francois and Magro, Marc and Vicedo, Benoit",
title          = "{An integrable deformation of the $AdS_5 \times S^5$ superstring action}",
journal        = "Phys. Rev. Lett.",
volume         = "112",
year           = "2014",
number         = "5",
pages          = "051601",
doi            = "10.1103/PhysRevLett.112.051601",
eprint         = "1309.5850",
archivePrefix  = "arXiv",
primaryClass   = "hep-th",
SLACcitation   = "
}

@article{Delduc:2014kha,
author         = "Delduc, Francois and Magro, Marc and Vicedo, Benoit",
title          = "{Derivation of the action and symmetries of the $q$-deformed $AdS_5 \times S^5$ superstring}",
journal        = "JHEP",
volume         = "10",
year           = "2014",
pages          = "132",
doi            = "10.1007/JHEP10(2014)132",
eprint         = "1406.6286",
archivePrefix  = "arXiv",
primaryClass   = "hep-th",
SLACcitation   = "
}

@article{Sorokin:2011rr,
author         = "Sorokin, Dmitri and Tseytlin, Arkady and Wulff, Linus and Zarembo, Konstantin",
title          = "{Superstrings in $AdS_2 \times S^2 \times T^6$}",
journal        = "J. Phys.",
volume         = "A44",
year           = "2011",
pages          = "275401",
doi            = "10.1088/1751-8113/44/27/275401",
eprint         = "1104.1793",
archivePrefix  = "arXiv",
primaryClass   = "hep-th",
reportNumber   = "MIFPA-11-11, NORDITA-2011-30, IMPERIAL-TP-AT-2011-2",
SLACcitation   = "
}

@article{Hoare:2016ibq,
author         = "Hoare, Ben and van Tongeren, Stijn J.",
title          = "{Non-split and split deformations of $AdS_5$}",
journal        = "J. Phys.",
volume         = "A49",
year           = "2016",
number         = "48",
pages          = "484003",
doi            = "10.1088/1751-8113/49/48/484003",
eprint         = "1605.03552",
archivePrefix  = "arXiv",
primaryClass   = "hep-th",
SLACcitation   = "
}

@article{Araujo:2018rbc,
author         = "Araujo, Thiago and Colg\'{a}in, Eoin \'{O}. and Yavartanoo, Hossein",
title          = "{Embedding the modified CYBE in Supergravity}",
journal        = "Eur. Phys. J.",
volume         = "C78",
year           = "2018",
number         = "10",
pages          = "854",
doi            = "10.1140/epjc/s10052-018-6335-6",
eprint         = "1806.02602",
archivePrefix  = "arXiv",
primaryClass   = "hep-th",
reportNumber   = "APCTP Pre2018-004, APCTP-PRE2018-004",
SLACcitation   = "
}

@article{Arutyunov:2009ga,
author         = "Arutyunov, Gleb and Frolov, Sergey",
title          = "{Foundations of the $AdS_5 \times S^5$ Superstring. Part I}",
journal        = "J. Phys.",
volume         = "A42",
year           = "2009",
pages          = "254003",
doi            = "10.1088/1751-8113/42/25/254003",
eprint         = "0901.4937",
archivePrefix  = "arXiv",
primaryClass   = "hep-th",
reportNumber   = "ITP-UU-09-05, SPIN-09-05, TCD-MATH-09-06, HMI-09-03",
SLACcitation   = "
}

@article{Grigoriev:2007bu,
author         = "Grigoriev, M. and Tseytlin, Arkady A.",
title          = "{Pohlmeyer reduction of $AdS_5 \times S^5$ superstring sigma model}",
journal        = "Nucl. Phys.",
volume         = "B800",
year           = "2008",
pages          = "450-501",
doi            = "10.1016/j.nuclphysb.2008.01.006",
eprint         = "0711.0155",
archivePrefix  = "arXiv",
primaryClass   = "hep-th",
reportNumber   = "IMPERIAL-TP-AT-2007-4",
SLACcitation   = "
}

@article{Hoare:2015gda,
author         = "Hoare, B. and Tseytlin, A. A.",
title          = "{On integrable deformations of superstring sigma models related to $AdS_n \times S^n$ supercosets}",
journal        = "Nucl. Phys.",
volume         = "B897",
year           = "2015",
pages          = "448-478",
doi            = "10.1016/j.nuclphysb.2015.06.001",
eprint         = "1504.07213",
archivePrefix  = "arXiv",
primaryClass   = "hep-th",
reportNumber   = "IMPERIAL-TP-AT-2015-02, HU-EP-15-21",
SLACcitation   = "
}

@article{Klimcik:2002zj,
author         = "Klim\v{c}\'{i}k, Ctirad",
title          = "{Yang-Baxter $\sigma$-models and dS/AdS T-duality}",
journal        = "JHEP",
volume         = "12",
year           = "2002",
pages          = "051",
doi            = "10.1088/1126-6708/2002/12/051",
eprint         = "hep-th/0210095",
archivePrefix  = "arXiv",
primaryClass   = "hep-th",
reportNumber   = "IML-02-XY",
SLACcitation   = "
}

@article{Delduc:2013fga,
author         = "Delduc, Francois and Magro, Marc and Vicedo, Benoit",
title          = "{On classical $q$-deformations of integrable $\sigma$-models}",
journal        = "JHEP",
volume         = "11",
year           = "2013",
pages          = "192",
doi            = "10.1007/JHEP11(2013)192",
eprint         = "1308.3581",
archivePrefix  = "arXiv",
primaryClass   = "hep-th",
SLACcitation   = "
}

@article{Bena:2003wd,
author         = "Bena, Iosif and Polchinski, Joseph and Roiban, Radu",
title          = "{Hidden symmetries of the $AdS_5 \times S^5$ superstring}",
journal        = "Phys. Rev.",
volume         = "D69",
year           = "2004",
pages          = "046002",
doi            = "10.1103/PhysRevD.69.046002",
eprint         = "hep-th/0305116",
archivePrefix  = "arXiv",
primaryClass   = "hep-th",
reportNumber   = "NSF-KITP-03-34, UCLA-03-TEP-14",
SLACcitation   = "
}

@article{Magro:2008dv,
author         = "Magro, Marc",
title          = "{The Classical Exchange Algebra of $AdS_5 \times S^5$}",
journal        = "JHEP",
volume         = "01",
year           = "2009",
pages          = "021",
doi            = "10.1088/1126-6708/2009/01/021",
eprint         = "0810.4136",
archivePrefix  = "arXiv",
primaryClass   = "hep-th",
reportNumber   = "AEI-2008-085",
SLACcitation   = "
}

@article{Vicedo:2010qd,
author         = "Vicedo, Benoit",
title          = "{The classical R-matrix of AdS/CFT and its Lie dialgebra structure}",
journal        = "Lett. Math. Phys.",
volume         = "95",
year           = "2011",
pages          = "249-274",
doi            = "10.1007/s11005-010-0446-9",
eprint         = "1003.1192",
archivePrefix  = "arXiv",
primaryClass   = "hep-th",
reportNumber   = "IPHT-T10-026",
SLACcitation   = "
}

@article{Severa:2018pag,
author         = "\v{S}evera, Pavol and Valach, Fridrich",
title          = "{Courant algebroids, Poisson-Lie T-duality, and type II supergravities}",
year           = "2018",
eprint         = "1810.07763",
archivePrefix  = "arXiv",
primaryClass   = "math.DG",
SLACcitation   = "
}

@article{Demulder:2018lmj,
author         = "Demulder, Saskia and Hassler, Falk and Thompson, Daniel C.",
title          = "{Doubled aspects of generalised dualities and integrable deformations}",
year           = "2018",
eprint         = "1810.11446",
archivePrefix  = "arXiv",
primaryClass   = "hep-th",
SLACcitation   = "
}

@article{Araujo:2017enj,
author         = "Araujo, T. and Colg\'{a}in, E. \'{O} and Sakamoto, J. and Sheikh-Jabbari, M. M. and Yoshida, K.",
title          = "{$I$ in generalized supergravity}",
journal        = "Eur. Phys. J.",
volume         = "C77",
year           = "2017",
number         = "11",
pages          = "739",
doi            = "10.1140/epjc/s10052-017-5316-5",
eprint         = "1708.03163",
archivePrefix  = "arXiv",
primaryClass   = "hep-th",
reportNumber   = "APCTP-PRE2017---015, KUNS-2696, IPM-P-2017-024,IPM-P-2017-028",
SLACcitation   = "
}

@article{Hoare:2014oua,
author         = "Hoare, Ben",
title          = "{Towards a two-parameter q-deformation of $AdS_3 \times S^3 \times M^4$ superstrings}",
journal        = "Nucl. Phys.",
volume         = "B891",
year           = "2015",
pages          = "259-295",
doi            = "10.1016/j.nuclphysb.2014.12.012",
eprint         = "1411.1266",
archivePrefix  = "arXiv",
primaryClass   = "hep-th",
reportNumber   = "HU-EP-14-44",
SLACcitation   = "
}

@article{Delduc:2018xug,
author         = "Delduc, Francois and Hoare, Ben and Kameyama, Takashi and Lacroix, Sylvain and Magro, Marc",
title          = "{Three-parameter integrable deformation of $\Integer_4$ permutation supercosets}",
year           = "2018",
eprint         = "1811.00453",
archivePrefix  = "arXiv",
primaryClass   = "hep-th",
reportNumber   = "ZMP-HH/18-22",
SLACcitation   = "
}

@article{Cagnazzo:2012se,
author         = "Cagnazzo, A. and Zarembo, K.",
title          = "{B-field in $AdS_3/CFT_2$ Correspondence and Integrability}",
journal        = "JHEP",
volume         = "11",
year           = "2012",
pages          = "133",
doi            = "10.1007/JHEP11(2012)133",
note           = "[Erratum: \doiref{10.1007/JHEP04(2013)003}{\textsf{JHEP 1304, 003 (2013)}}]",
eprint         = "1209.4049",
archivePrefix  = "arXiv",
primaryClass   = "hep-th",
reportNumber   = "NORDITA-2012-67, UUITP-24-12",
SLACcitation   = "
}

@article{Hoare:2018ebg,
author         = "Hoare, Ben and Seibold, Fiona K.",
title          = "{Poisson-Lie duals of the $\eta$-deformed $AdS_2 \times S^2 \times T^6$ superstring}",
journal        = "JHEP",
volume         = "08",
year           = "2018",
pages          = "107",
doi            = "10.1007/JHEP08(2018)107",
eprint         = "1807.04608",
archivePrefix  = "arXiv",
primaryClass   = "hep-th",
SLACcitation   = "
}

@article{Alvarez:1994np,
author         = "\'{A}lvarez, Enrique and \'{A}lvarez-Gaum\'{e}, Luis and Lozano, Yolanda",
title          = "{On non-abelian duality}",
journal        = "Nucl. Phys.",
volume         = "B424",
year           = "1994",
pages          = "155-183",
doi            = "10.1016/0550-3213(94)90093-0",
eprint         = "hep-th/9403155",
archivePrefix  = "arXiv",
primaryClass   = "hep-th",
reportNumber   = "CERN-TH-7204-94",
SLACcitation   = "
}

@article{Elitzur:1994ri,
author         = "Elitzur, S. and Giveon, A. and Rabinovici, E. and Schwimmer, A. and Veneziano, G.",
title          = "{Remarks on non-abelian duality}",
journal        = "Nucl. Phys.",
volume         = "B435",
year           = "1995",
pages          = "147-171",
doi            = "10.1016/0550-3213(94)00426-F",
eprint         = "hep-th/9409011",
archivePrefix  = "arXiv",
primaryClass   = "hep-th",
reportNumber   = "CERN-TH-7414-94, RI-9-94, WIS-7-94",
SLACcitation   = "
}

@article{Tyurin:1995bu,
author         = "Tyurin, Eugene and von Unge, Rikard",
title          = "{Poisson-Lie T-duality: the path-integral derivation}",
journal        = "Phys. Lett.",
volume         = "B382",
year           = "1996",
pages          = "233-240",
doi            = "10.1016/0370-2693(96)00680-6",
eprint         = "hep-th/9512025",
archivePrefix  = "arXiv",
primaryClass   = "hep-th",
reportNumber   = "ITP-SB-95-50, USITP-95-11",
SLACcitation   = "
}

@article{Bossard:2001au,
author         = "Bossard, A. and Mohammedi, N.",
title          = "{Poisson-Lie duality in the string effective action}",
journal        = "Nucl. Phys.",
volume         = "B619",
year           = "2001",
pages          = "128-154",
doi            = "10.1016/S0550-3213(01)00541-7",
eprint         = "hep-th/0106211",
archivePrefix  = "arXiv",
primaryClass   = "hep-th",
SLACcitation   = "
}

@article{Zarembo:2010sg,
author         = "Zarembo, K.",
title          = "{Strings on semisymmetric superspaces}",
journal        = "JHEP",
volume         = "05",
year           = "2010",
pages          = "002",
doi            = "10.1007/JHEP05(2010)002",
eprint         = "1003.0465",
archivePrefix  = "arXiv",
primaryClass   = "hep-th",
reportNumber   = "ITEP-TH-12-10, LPTENS-10-12, UUITP-05-10",
SLACcitation   = "
}

@article{Fateev:1992tk,
author         = "Fateev, V. A. and Onofri, E. and Zamolodchikov, Alexei B.",
title          = "{The sausage model (integrable deformations of O(3) sigma model)}",
journal        = "Nucl. Phys.",
volume         = "B406",
year           = "1993",
pages          = "521-565",
doi            = "10.1016/0550-3213(93)90001-6",
reportNumber   = "PAR-LPTHE-92-46, LPTHE-92-46",
SLACcitation   = "
}

@article{Klimcik:2008eq,
author         = "Klim\v{c}\'{i}k, Ctirad",
title          = "{On integrability of the Yang-Baxter $\sigma$-model}",
journal        = "J. Math. Phys.",
volume         = "50",
year           = "2009",
pages          = "043508",
doi            = "10.1063/1.3116242",
eprint         = "0802.3518",
archivePrefix  = "arXiv",
primaryClass   = "hep-th",
SLACcitation   = "
}

@article{Klimcik:2014bta,
author         = "Klim\v{c}\'{i}k, Ctirad",
title          = "{Integrability of the bi-Yang-Baxter sigma-model}",
journal        = "Lett. Math. Phys.",
volume         = "104",
year           = "2014",
pages          = "1095-1106",
doi            = "10.1007/s11005-014-0709-y",
eprint         = "1402.2105",
archivePrefix  = "arXiv",
primaryClass   = "math-ph",
SLACcitation   = "
}

@article{Drinfeld:1985rx,
author         = "Drinfel'd, V. G.",
title          = "{Hopf algebras and the quantum Yang-Baxter equation}",
journal        = "Sov. Math. Dokl.",
volume         = "32",
year           = "1985",
pages          = "254",
note           = "[\textsf{Dokl. Akad. Nauk Ser. Fiz. 283, 1060 (1985)}]",
SLACcitation   = "
}

@article{Jimbo:1985zk,
author         = "Jimbo, Michio",
title          = "{A q-difference analog of U(g) and the Yang-Baxter equation}",
journal        = "Lett. Math. Phys.",
volume         = "10",
year           = "1985",
pages          = "63",
doi            = "10.1007/BF00704588",
SLACcitation   = "
}

@article{Belavin:1984,
author         = "Belavin, A. A. and Drinfel'd, V. G.",
title          = "{Triangle equations and simple Lie algebras}",
journal        = "Sov. Sci. Rev.",
volume         = "C4",
year           = "1984",
pages          = "93",
}

@article{SemenovTianShansky:1983ik,
author         = "Semenov-Tian-Shansky, M. A.",
title          = "{What is a classical r-matrix?}",
journal        = "Funct. Anal. Appl.",
volume         = "17",
year           = "1983",
pages          = "259-272",
doi            = "10.1007/BF01076717",
note           = "[\textsf{Funkt. Anal. Pril. 17, 17 (1983)}]",
SLACcitation   = "
}

@article{Klimcik:1995ux,
author         = "Klim\v{c}\'{i}k, C. and \v{S}evera, P.",
title          = "{Dual non-abelian duality and the Drinfel'd double}",
journal        = "Phys. Lett.",
volume         = "B351",
year           = "1995",
pages          = "455-462",
doi            = "10.1016/0370-2693(95)00451-P",
eprint         = "hep-th/9502122",
archivePrefix  = "arXiv",
primaryClass   = "hep-th",
reportNumber   = "CERN-TH-95-39, CERN-TH-95-039",
SLACcitation   = "
}

@article{Klimcik:1995jn,
author         = "Klim\v{c}\'{i}k, C.",
title          = "{Poisson-Lie T-duality}",
journal        = "Nucl. Phys. Proc. Suppl.",
volume         = "46",
year           = "1996",
pages          = "116-121",
doi            = "10.1016/0920-5632(96)00013-8",
eprint         = "hep-th/9509095",
archivePrefix  = "arXiv",
primaryClass   = "hep-th",
reportNumber   = "CERN-TH-95-248",
SLACcitation   = "
}

@article{Klimcik:1995dy,
author         = "Klim\v{c}\'{i}k, C. and \v{S}evera, P.",
title          = "{Poisson-Lie T-duality and loop groups of Drinfel'd doubles}",
journal        = "Phys. Lett.",
volume         = "B372",
year           = "1996",
pages          = "65-71",
doi            = "10.1016/0370-2693(96)00025-1",
eprint         = "hep-th/9512040",
archivePrefix  = "arXiv",
primaryClass   = "hep-th",
reportNumber   = "CERN-TH-95-330",
SLACcitation   = "
}

@article{Klimcik:1996nq,
author         = "Klim\v{c}\'{i}k, C. and \v{S}evera, P.",
title          = "{Non-abelian momentum-winding exchange}",
journal        = "Phys. Lett.",
volume         = "B383",
year           = "1996",
pages          = "281-286",
doi            = "10.1016/0370-2693(96)00755-1",
eprint         = "hep-th/9605212",
archivePrefix  = "arXiv",
primaryClass   = "hep-th",
reportNumber   = "CERN-TH-96-142",
SLACcitation   = "
}

@article{Delduc:2016ihq,
author         = "Delduc, Francois and Lacroix, Sylvain and Magro, Marc and Vicedo, Benoit",
title          = "{On $q$-deformed symmetries as Poisson-Lie symmetries and application to Yang-Baxter type models}",
journal        = "J. Phys.",
volume         = "A49",
year           = "2016",
number         = "41",
pages          = "415402",
doi            = "10.1088/1751-8113/49/41/415402",
eprint         = "1606.01712",
archivePrefix  = "arXiv",
primaryClass   = "hep-th",
SLACcitation   = "
}

@article{Green:1983wt,
author         = "Green, Michael B. and Schwarz, John H.",
title          = "{Covariant Description of Superstrings}",
journal        = "Phys. Lett.",
volume         = "B136",
year           = "1984",
pages          = "367",
doi            = "10.1016/0370-2693(84)92021-5",
reportNumber   = "QMC-83-7",
SLACcitation   = "
}

@article{Grisaru:1985fv,
author         = "Grisaru, Marcus T. and Howe, Paul S. and Mezincescu, L. and Nilsson, B. and Townsend, P. K.",
title          = "{N=2 Superstrings in a Supergravity Background}",
journal        = "Phys. Lett.",
volume         = "B162",
year           = "1985",
pages          = "116",
doi            = "10.1016/0370-2693(85)91071-8",
reportNumber   = "Print-85-0603 (CAMBRIDGE)",
SLACcitation   = "
}

@article{Tseytlin:1996hs,
author         = "Tseytlin, Arkady A.",
title          = "{On dilaton dependence of type II superstring action}",
journal        = "Class. Quant. Grav.",
volume         = "13",
year           = "1996",
pages          = "L81",
doi            = "10.1088/0264-9381/13/6/003",
eprint         = "hep-th/9601109",
archivePrefix  = "arXiv",
primaryClass   = "hep-th",
reportNumber   = "IMPERIAL-TP-95-96-19",
SLACcitation   = "
}

@article{Cvetic:1999zs,
author         = "Cveti\v{c}, Mirjam and Lu, Hong and Pope, C. N. and Stelle, K. S.",
title          = "{T-duality in the Green-Schwarz formalism, and the massless/massive IIA duality map}",
journal        = "Nucl. Phys.",
volume         = "B573",
year           = "2000",
pages          = "149",
doi            = "10.1016/S0550-3213(99)00740-3",
eprint         = "hep-th/9907202",
archivePrefix  = "arXiv",
primaryClass   = "hep-th",
reportNumber   = "UPR-0852-T, CTP-TAMU-31-99, SISSA-88-99-EP,
IMPERIAL-TP-98-99-63, NSF-ITP-99-086",
SLACcitation   = "
}

@article{Wulff:2013kga,
author         = "Wulff, Linus",
title          = "{The type II superstring to order $\theta^4$}",
journal        = "JHEP",
volume         = "07",
year           = "2013",
pages          = "123",
doi            = "10.1007/JHEP07(2013)123",
eprint         = "1304.6422",
archivePrefix  = "arXiv",
primaryClass   = "hep-th",
reportNumber   = "MIFPA-13-13",
SLACcitation   = "
}

@article{Borowiec:2015wua,
author         = "Borowiec, Andrzej and Kyono, Hideki and Lukierski, Jerzy and Sakamoto, Jun-ichi and Yoshida, Kentaroh",
title          = "{Yang-Baxter sigma models and Lax pairs arising from $\kappa$-Poincar\'{e} $r$-matrices}",
journal        = "JHEP",
volume         = "04",
year           = "2016",
pages          = "079",
doi            = "10.1007/JHEP04(2016)079",
eprint         = "1510.03083",
archivePrefix  = "arXiv",
primaryClass   = "hep-th",
reportNumber   = "KUNS-2578",
SLACcitation   = "
}

@article{Wulff:2014kja,
author         = "Wulff, Linus",
title          = "{Superisometries and integrability of superstrings}",
journal        = "JHEP",
volume         = "05",
year           = "2014",
pages          = "115",
doi            = "10.1007/JHEP05(2014)115",
eprint         = "1402.3122",
archivePrefix  = "arXiv",
primaryClass   = "hep-th",
reportNumber   = "IMPERIAL-TP-LW-2014-01",
SLACcitation   = "
}

@article{Borsato:2018idb,
author         = "Borsato, Riccardo and Wulff, Linus",
title          = "{Non-abelian T-duality and Yang-Baxter deformations of Green-Schwarz strings}",
journal        = "JHEP",
volume         = "08",
year           = "2018",
pages          = "027",
doi            = "10.1007/JHEP08(2018)027",
eprint         = "1806.04083",
archivePrefix  = "arXiv",
primaryClass   = "hep-th",
reportNumber   = "NORDITA 2018-041, NORDITA-2018-041",
SLACcitation   = "
}

@article{Roychowdhury:2018qsz,
author         = "Roychowdhury, Dibakar",
title          = "{On pp wave limit for $\eta$ deformed superstrings}",
journal        = "JHEP",
volume         = "05",
year           = "2018",
pages          = "018",
doi            = "10.1007/JHEP05(2018)018",
eprint         = "1801.07680",
archivePrefix  = "arXiv",
primaryClass   = "hep-th",
SLACcitation   = "
}

@article{Berenstein:2002jq,
author         = "Berenstein, David Eliecer and Maldacena, Juan Martin and Nastase, Horatiu Stefan",
title          = "{Strings in flat space and pp waves from $\mathcal{N} = 4$ Super Yang Mills}",
journal        = "JHEP",
volume         = "04",
year           = "2002",
pages          = "013",
doi            = "10.1088/1126-6708/2002/04/013",
eprint         = "hep-th/0202021",
archivePrefix  = "arXiv",
primaryClass   = "hep-th",
SLACcitation   = "
}

@article{Blau:2002dy,
author         = "Blau, Matthias and Figueroa-O'Farrill, Jose M. and Hull, Christopher and Papadopoulos, George",
title          = "{Penrose limits and maximal supersymmetry}",
journal        = "Class. Quant. Grav.",
volume         = "19",
year           = "2002",
pages          = "L87-L95",
doi            = "10.1088/0264-9381/19/10/101",
eprint         = "hep-th/0201081",
archivePrefix  = "arXiv",
primaryClass   = "hep-th",
reportNumber   = "EMPG-02-01, QMUL-PH-02-01",
SLACcitation   = "
}

@article{Sakamoto:2017wor,
author         = "Sakamoto, Jun-ichi and Sakatani, Yuho and Yoshida, Kentaroh",
title          = "{Weyl invariance for generalized supergravity backgrounds from the doubled formalism}",
journal        = "PTEP",
volume         = "2017",
year           = "2017",
number         = "5",
pages          = "053B07",
doi            = "10.1093/ptep/ptx067",
eprint         = "1703.09213",
archivePrefix  = "arXiv",
primaryClass   = "hep-th",
reportNumber   = "KUNS-2668",
SLACcitation   = "
}

@article{Fernandez-Melgarejo:2018wpg,
author         = "Fern\'{a}ndez-Melgarejo, Jose J. and Sakamoto, Jun-ichi and Sakatani, Yuho and Yoshida, Kentaroh",
title          = "{Comments on Weyl invariance of string theories in generalized supergravity backgrounds}",
year           = "2018",
eprint         = "1811.10600",
archivePrefix  = "arXiv",
primaryClass   = "hep-th",
SLACcitation   = "
}

\end{bibtex}

\bibliographystyle{nb}
\bibliography{\jobname}

\end{document}